\documentclass[10pt]{iopart}
\usepackage{graphicx,caption}
\usepackage{hyperref}

\usepackage{wrapfig}

\usepackage{iopams}
\begin{document}

\title[]{Polarization and phase control of electron injection and acceleration in the plasma by a self-steepening laser pulse}

\author{Jihoon Kim$^1$, Tianhong Wang$^1$, Vladimir Khudik$^2$, Gennady Shvets$^1$}

\address{$^1$School of Applied and Engineering Physics, Cornell University, Ithaca, NY 14850, USA.\\$^2$Department of Physics and Institute for Fusion Studies, The University of Texas at Austin, Austin, TX 78712, USA.}
\ead{jk2628@cornell.edu, gshvets@cornell.edu}
\vspace{10pt}
\begin{indented}
\item[]Oct 2022
\end{indented}

\begin{abstract}
We describe an interplay between two injection mechanism of background electrons into an evolving plasma bubble behind an intense laser pulse: one due to the overall bubble expansion, and another due to its periodic undulation. The two mechanisms occur simultaneously when an intense laser pulse propagating inside a plasma forms a shock-like steepened front. Periodic undulations of the plasma bubble along the laser propagation path can either inhibit or conspire with electron injection due to bubble expansion. We show that Carrier-Envelope-Phase (CEP) controlled plasma bubble undulation induced by the self-steepening laser pulse produces a unique electron injector -- Expanding Phase-controlled Undulating Bubble (EPUB).  The longitudinal structure of the electron bunch injected by the EPUB can be controlled by laser polarization and power, resulting in high-charge (multiple nano-Coulombs) high-current (tens of kilo-Amperes) electron beams with ultra-short (femtosecond-scale) temporal structure. Generation of high-energy betatron radiation with polarization- and CEP-controlled energy spectrum and angular distribution is analyzed as a promising application of EPUB-produced beams.
\end{abstract}

%
%
%
%
%

\section{Introduction}

An electron injector is an integral part of any accelerator, as it produces high-quality moderate energy particles for further acceleration. A remarkable feature of a Laser Wakefield Accelerator(LWFA)~\cite{Malka,Hooker,RMPS} is the availability of an abundant reservoir of charged particles from the background plasma. Therefore, plasma can simultaneously serve as an acceleration medium sustaining intense plasma waves, and an electron injector. While the key attraction of LWFAs is their compactness owing to ultrahigh accelerating electric field -- in excess of $100$GV/m in many recent implementations~\cite{Nakamura,downer_nat_comm,Lu,Leemans,Kim,Gonsalves} -- of the plasma wave generated by intense laser pulses, its other advantage is the availability  of large numbers of initially quiescent electrons that can be injected into the plasma wave, capable of forming currents exceeding 100kA~\cite{170kA}. If such an injection can be controlled, it may be possible to produce high-charge low-emittance beams in single compact device.

A number of promising approaches to injecting electrons into plasma waves generated in the wake of a laser pulse, including the highly nonlinear ``plasma bubbles" ~\cite{kostyukov_pop04,rosenzweig}, have been suggested and experimentally implemented. Those include injections due to ionization~\cite{chen_jap06,pak_prl10,McGuffey_prl10,mori_bunches_prl16,Min_Chen_CEP}, engineered density ramps~\cite{geddes_prl08,schmid_prstab10,buck_prl13,gonsalves_nphys11}, and rapid variation of the bubble's size along the laser's path~\cite{kalmykov_prl,austin_ppcf,downer_nat_comm,austin_pop13,pak,suk_prl,malka_prl13}.

Electron injection and acceleration based on single-cycle laser pulses has been demonstrated theoretically and experimentally ~\cite{kost_cep,CEP_observable,Zhengyan,kHz_injection,Veisz_2cycle,Salehi,Jihoon}. Under certain circumstances, near single cycle (NSC) laser pulse propagating in an underdense plasma can generate a phase-controlled undulating bubble (PUB) with characteristic periodicity $T_{\rm CEP}=\lambda_L/(v_{\rm ph}-v_{\rm g})$ controlled by laser Carrier Envelope Phase (CEP) offset, with $v_{\rm ph}$ and $v_{\rm g}$, the laser phase and group velocity ~\cite{CEP_observable,Zhengyan,Salehi,kost_cep}. CEP-controlled injection is expected when laser intensity varies sharply on a time-scale of one laser oscillation period -- either due to the short overall duration, or nonlinear self-steepening of a laser pulse in the course of its propagation through the plasma~\cite{ma_screp16,Vieira_steepen,Decker_steepen,Najmudin_steepen,krush_prl14}.

It was recently shown that CEP-related periodic electron injection into a plasma bubble can occur for intense NSC laser pulses~\cite{Jihoon}. This CEP-controlled injection is a conceptual departure from the standard description of plasma wave generation by multi-cycle laser pulses that relies on the phase-averaged (ponderomotive) approximation~\cite{Mora}. Despite the promise of CEP-based injection to generating high-current ultra-short electron bunches~\cite{Jihoon}, it requires NSC pulses. In what follows, we concentrate on the other circumstance under which phase- and polarization-dependent injection can occur: when a longer pulse has its front locally depleted due to the etching by the plasma. The front of such self-steepened laser pulse envelope can vary on a scale comparable to that of a laser cycle~\cite{seidel_arxiv22,ma_screp16,Vieira_steepen,Decker_steepen}, resulting in an Expanding Phase-controlled Undulating Bubble (EPUB) which is the subject of this work.

\begin{figure}[t]
    \includegraphics[width=\textwidth]{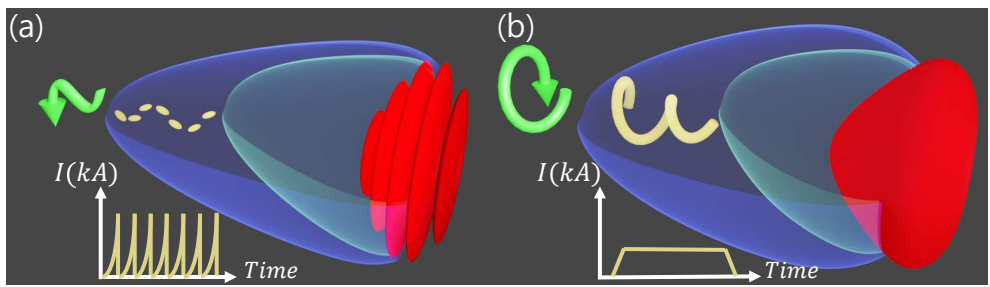}
\caption{Schematic of Injection process for plasma bubble driven by (a) linearly polarized laser pulse  and (b) circularly polarized laser pulse. An expanding and transversely undulating bubble is formed by a laser with steepened front. According to laser polarization, periodically modulated or flat current beam can be generated. Laser intensity (red), plasma bubble shape at early (light blue) and later (dark blue) time, electron bunch density (yellow),  plasma bubble back movement (green arrows) and injected current (white axes, yellow lines)}\label{fig:schematic}
\end{figure}

In this paper, we examine the combined effect of expansion and undulation of a plasma bubble on the injection, acceleration, and temporal shaping of an electron bunch produced by an EPUB, as shown in Fig \ref{fig:schematic}.  The paper is organized as follows. In Section~\ref{sec:pic}, we set the stage by presenting the results of PIC simulations that demonstrate phase and polarization dependent injection of electrons into a plasma bubble produced by a self steepening few cycle laser pulse with $cT_{\rm FWHM} \sim 3 \lambda_L$, where $T_{\rm FWHM}$ and $\lambda_L$ are the pulse duration and wavelength, respectively. The parameters of the laser pulse are chosen to be within reach of the BESTIA (Brookhaven Experimental Supra-Terawatt Infrared at ATF) laser system~\cite{BESTIA}. In Section~\ref{sec:analytics}, we interpret these results by developing a simple single-particle model of electron injection into a plasma bubble undergoing simultaneous expansion and undulation. This model is used to demonstrate how laser polarization (i.e. linear versus circular) can be used to generate the desired current profile (spiky versus smooth) of an injected electron bunch, and further lead to X-ray distribution with asymmetric angular distribution and nonzero degree of polarization. We demonstrate that high charge ($Q\sim 10 {\rm nC}$) bunches modulated on a temporal scale comparable to the laser period $T_L = \lambda_L/c$ can be formed, with promising implications for structured X-ray generation. In Section ~\ref{section:discussions}, we discuss the laser-to-bunch energy transfer efficiency of the proposed injection/acceleration scheme, the possibility of controlling the spectrum and angular distribution of the betatron radiation using laser polarization and CEP, and the possibility of extracting telltale signs of phase-linked laser-plasma interactions using betatron radiation.

\section{Polarization-dependent injection and acceleration: simulations results}\label{sec:pic}

We use a 3D PIC code VLPL~\cite{Pukhov_code} to self-consistently model the propagation and self-steepening of an intense laser pulse~\cite{Vieira_steepen, Decker_steepen}, followed by self-injection of some of the plasma electrons into the laser wakefield, acceleration of the injected bunch, and subsequent plasma field depletion by the injected electrons~\cite{beamloading}. The following laser parameters are used: peak power $P_L = 40 {\rm TW}$, wavelength $\lambda_L= 9.2 {\rm \mu m}$, pulse duration $T_{\rm FWHM}=100 {\rm fs}$, and the matched spot size $\sigma_\perp=8.5\lambda_L \approx 78 {\rm \mu m}$. This corresponds to the initial normalized vector potential of the laser pulse $a_{L} \equiv eE_{\perp}/mc\omega_L= 5.0$, where $E_{\perp}$ is the peak electric field of the laser pulse and $\omega_L \equiv 2\pi c/\lambda_L$ is its angular frequancy. Pre-ionized plasma is assumed to start with a linear density ramp of the length $L_{\rm ramp} = 0.37 {\rm mm}$, followed by a long plateau region with constant density $n=9.1\times 10^{16}/cm^3$. A numerical grid used in the simulations was chosen to have the dimensions of $\Delta x \times \Delta y \times \Delta z = 0.05\lambda_L \times 0.25 \lambda_L \times 0.25 \lambda_L$ and a time step $\Delta t=0.05\lambda_L/c$, where $x$ is the propagation direction of the laser pulse through the plasma (Also see Table \ref{table1}). We note that polarization-dependent electron injection has been observed in initially-neutral plasma targets due to above-threshold ionization (ATI) process~\cite{ma_dollar_prl20}. Ionization injection is neglected in our simulations because the total injected charge ($Q > 10 {\rm nC}$) due to plasma bubble expansion/undulation is expected to be much larger than in the case of ionization injection~\cite{ma_dollar_pop21}.

\begin{table}[ht]
\caption{\label{table1}Laser-Plasma parameters used in a 3D PIC (VLPL) simulation. Laser field is defined by the normalized vector potential $ a(x,y,z) = - a_0 \exp \{-(z^2+y^2)/\sigma_\perp^2\}\exp ({-{x^2}/\sigma_x^2}) \cos\{{\omega_L (x/c- t)+\phi_{CEP}}\} $ and $P_c\equiv17\omega_L^2/\omega_p^2 \rm{(GW)}$\cite{RMPS} }
\begin{tabular*}{\textwidth}{@{}l*{15}{@{\extracolsep{0pt plus
12pt}}l}}
\br
Parameter & Physical Units & Normalized Units \\
\mr
Plasma Density &$9.1\times 10^{16}/cm^3$ & $n_p/n_c=1/144$ \\
Ramp Length & 0.37mm &$40\lambda_L$ \\ Cell Size ($\Delta x\times \Delta y\times \Delta z$) &  $0.46\mu m\times 2.3\mu m \times 2.3\mu m$ & $0.05\lambda_L\times 0.25\lambda_L \times 0.25\lambda_L$ \\ Plasma Wavelength($\lambda_p$) & $110\mu$m &12 $\lambda_L$ \\ 
Laser Wavelength($\lambda_L$) & 9.2 $\mu$m & $\lambda_L$\\ Spot Size $(\sigma_\perp)$ & 78$\mu$m &8.5$\lambda_L$  \\  FWHM ($\sqrt{2 \ln{2}}\sigma_x/c$)  & 100fs & 3.3 $\sigma_x/c$ \\ $a_0\equiv e|E_{\perp}|_{\rm max}/m_e c\omega_L$ & 1.7TV/m & 5.0 \\ Peak Power &40TW & $P/P_c=16.3$ \\
\br
\end{tabular*}
\end{table}

\begin{figure}[h]
    \includegraphics[width=\textwidth]{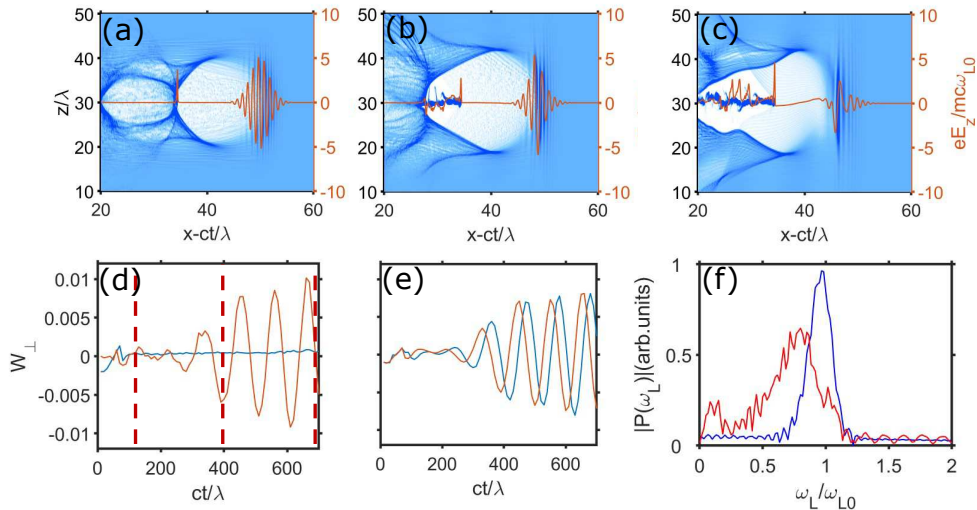}
\caption{Results of a three-dimensional particle-in-cell simulations of plasma bubble undulations induced by self-steepening of a laser pulse. (a-c) On-axis electric field of a $z$-polarized laser pulse (red line) and plasma density (color map) at (a) $ct=100\lambda_L$ , (b) $ct=400\lambda_L$, and (c) $ct=700\lambda_L$.  (d-e) On-axis transverse wakefield ${\bf W}_{\perp} = \left(W_y, W_z \right)$ for (d) linearly- and (e) circularly-polarized pulse. Blue (orange) line: $W_y$ ($W_z$) at $\xi \equiv x-ct = 35$. Red dashed lines on (d): propagation distances corresponding to (a), (b), and (c). (f) Spectra of the laser pulse at $ct=100\lambda_L$ (blue line) and at $ct=400\lambda_L$ (red line). Laser pulse parameters: peak power $P_L = 40 {\rm TW}$, wavelength $\lambda_L= 9.2 {\rm \mu m}$, duration $T_{\rm FWHM}=100 {\rm fs}$, spot size $\sigma_\perp=8.5\lambda_L \approx 78 {\rm \mu m}$. Plateau plasma density: $n=9.1\times 10^{16}/{\rm cm}^3$.}\label{fig:sim}
\end{figure}

We first consider a laser pulse linearly-polarized (LP) in the $z-$direction. Since the pulse front needs to steepen before CEP effect becomes visible, the plasma bubble does not execute transverse undulations immediately after the laser pulse enters the plasma as shown in Fig~\ref{fig:sim} (a) and (d). Electrons are injected into the plasma bubble from the very beginning, but this initial population of injected electrons does not exhibit any transverse asymmetry in the $z-$direction. After $ct=400\lambda_L$ (or $x=3.7 {\rm mm}$) of propagation through the plasma, the pulse front is depleted and steepened as shown in Fig.~\ref{fig:sim}(b), with further depletion at $ct=700\lambda_L$(or $x=6.4 {\rm mm}$) as apparent from Fig.~\ref{fig:sim}(c).

The sharpness of the self-steepened front at $ct=400\lambda_L$, as well as its depletion, are reflected in its spectrum plotted (red line) in Fig.~\ref{fig:sim}(b). When compared with the initial laser spectrum at $ct=100\lambda_L$ (blue-line), the spectrum of the steepened pulse is red-shifted by approximately $25\%$, i.e. from $\omega_L = \omega_{L0}$ to $\omega_L = 0.75\omega_{L0}$ -- a clear evidence of pulse depletion via plasma wake generation [Fig \ref{fig:sim}(f)]. Moreover, its large FWHM spectral bandwidth $\Delta \omega \sim 0.5 \omega_L$
signifies pulse steepening on the time scale of a laser period. This increase of the spectral bandwidth is a result of strongly-nonlinear interaction between the laser pulse and the plasma.

When the spatial profile of a laser pulse is self-steepened by its propagation through the plasma so as to develop a wavelength-sharp intensity shock, the plasma bubble starts executing transverse undulations along the laser polarization direction~\cite{ma_screp16}. Such plasma bubble undulations are analogous to those produced by NSC laser pulses~\cite{kost_cep,CEP_observable,Zhengyan,kHz_injection,Veisz_2cycle,Salehi,Jihoon}. Bubble undulations are manifested as a non-vanishing transverse on-axis  wakefield ${\bf W}_{\perp} \equiv \left(W_y, W_z \right)$. In its dimensionless form, the transverse wake is given by ${\bf W}_{\perp}  = \left( \mathbf{E}_{\perp} + \mathbf{e}_x \times \mathbf{B}_{\perp} \right)/E_{\rm 0L}$, where $\mathbf{E}_{\perp}$ and $\mathbf{B}_{\perp}$ are the transverse electric and magnetic wakefields inside the plasma bubble, and $E_{\rm 0L} = m_e c\omega_L/e$ is the relativistic electric field scale in a laser beam. Note that ${\bf W}_{\perp}$ is directly proportional to the transverse force exerted by the wakefield on a charge moving with the speed of light in the $x$-direction inside the plasma bubble.

When a laser pulse is linearly polarized (LP) in the $z-$direction, we find that $W_y \approx 0$ and $W_z \neq 0$ as shown in Fig.~\ref{fig:sim}(d). Similarly, an on-axis transverse wake ${\bf W}_{\perp}$ is generated by a circularly polarized (CP) laser pulse of the same duration and intensity as the LP pulse considered above. As evidenced by Fig.~\ref{fig:sim}(d)-(e), plasma cavity undulations begin around the same propagation distance for the LP and CP pulses because pulse steepening takes place after the same (polarization-independent) propagation distance through the plasma.

However, after bubble undulations start, their nature is observed to be highly dependent on the polarization state of the laser pulse. Specifically, instead of executing undulations in the $z-$ direction for the LP pulse, the bubble executes helical motion in the $y-z$ plane when driven by a CP pulse. As a result, the on-axis transverse wakefield $W_{\perp}$ of the bubble has equal $W_y$ and $W_z$ components that are approximately offset from each other by $\pi/2$ phase difference, as observed in Fig.~\ref{fig:sim}(e). As we are going to show in Section~\ref{sec:analytics}, it is important to note that for both polarizations, the plasma bubble is continuously elongated. Notably, the longitudinal bubble elongation is larger than its transverse expansion, as shown in Figs.~\ref{fig:sim}(a)-(c).

\begin{figure}[t]
    \includegraphics[width=\textwidth]{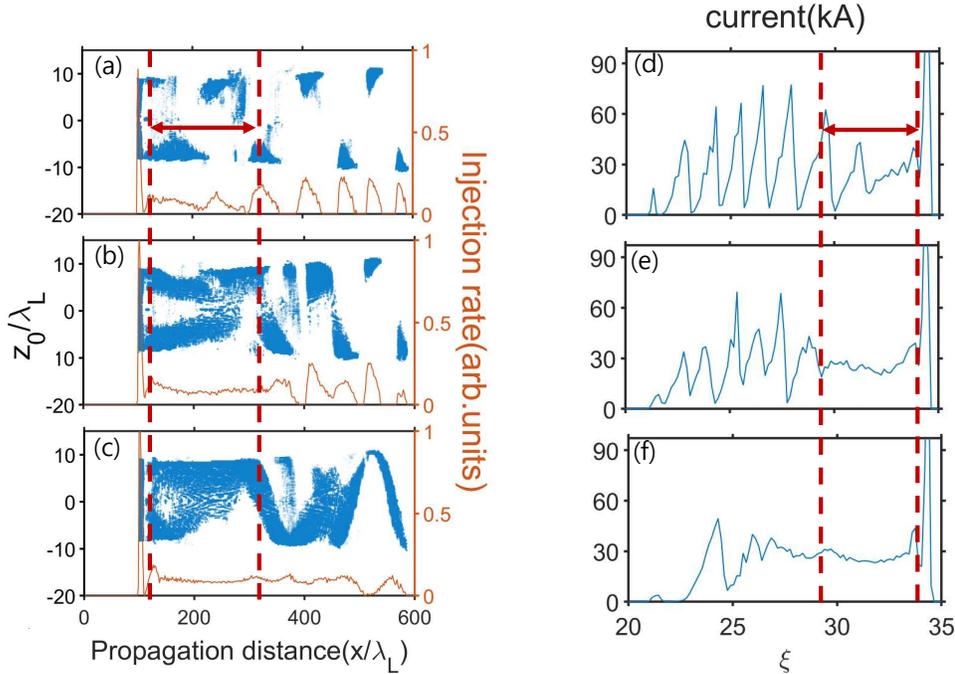}
\caption{Effect of laser ellipticity on injection and bunch formation. (a-c): Initial positions of the injected electrons in the $x_0-z_0$ plane (blue dots) and the injection rate (red line) for (a) linearly-polarized ($\epsilon=0$), (b) elliptically-polarized ($\epsilon=0.268$), and (c) circularly-polarized ($\epsilon=1$).  (d)-(f) Current profiles after $ct=6.4$mm propagation distance for (d) linear, (e) elliptic, and (f) circular laser polarizations. Red dashed lines: monoenergetic electrons, see Fig.~\ref{fig:spectra}. Ellipticity coefficients: $\epsilon = E_{{\bf L}y}/E_{{\bf L}z}$, where $E_{{\bf L}y,z}$ are laser electric field components. Laser and plasma parameters: same as in Fig.~\ref{fig:sim}.}\label{fig:injection}
\end{figure}

\subsection{Electron injection and its dependence on laser pulse polarizaton}\label{subsec:electron_groups}

For any laser polarization, copious amounts of electrons are injected from the background plasma throughout the entire laser propagation distance as shown in  Figs.~\ref{fig:injection}(a)-(c). For the specific laser and plasma parameters corresponding to the simuations in Figs.~\ref{fig:sim}-\ref{fig:injection}, the total injected charge is on the order of $Q_{\rm tot} \sim 11 {\rm nC}$ for all polarizations. However, the actual nature of electron injection is dependent on both the longitudinal injection location $x_0 \approx ct$ and the laser pulse polarization characterized by its ellipticity $\epsilon$ defined as $\epsilon = E_{{\bf L}y}/E_{{\bf L}z}$, where $E_{{\bf L}y,z}$ are the electric field components in an elliptically-polarized (EP) laser pulse. As the laser polarization is progressively varied from linear to elliptical to circular in Figs.~\ref{fig:injection} (d)-(f), current modulations switch from highly-bunched to weakly-modulated to nearly-constant.

In our simulation, every macro-particle is labeled by its initial location $\mathbf{r_0} \equiv \left( x_0, y_0, z_0 \right)$ that can be used to study the origin of injected electrons. Note that even though the longitudinal injection location $x_0$ (see Figs.~\ref{fig:injection}(a-c)) is approximately equal to time/distance expressed as $ct$ for a given snapshot of the electron density (see Fig.~\ref{fig:sim}(a-c) for three representative snapshots), these two quantities are not identical. For example, all injected electrons shown in a snapshot corresponding to the propagation distance $ct$ have been injected at earlier times corresponding to $x_0 \leq ct$.

\begin{figure}[h]
    \includegraphics[width=\textwidth]{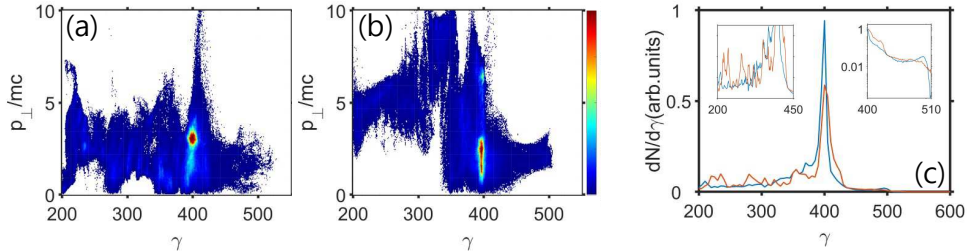}
\caption{Phase space distributions (a,b) and energy spectra (c) of the injected electrons at $ct = 6.4$mm for two laser pulse polarizations: LP (a) and CP (b). (c) Electron spectra for LP (red) and CP (blue) pulses. Low-energy inset: Group III electrons.  High-energy inset: Group II electrons.}\label{fig:spectra}
\end{figure}

Depending on the longitudinal injection location $x_0$, we have identified three groups of injected electrons: the earliest-injected during the laser passage through, and immediately after, the plasma density ramp (Group I); electrons injected during the passage through the $110\lambda_L < x_0 < 300\lambda_L$ region of the plasma marked by two vertical dashed lines in Fig.~\ref{fig:injection} that eventually form a monoenergetic bunch shown in Fig.~\ref{fig:spectra} (Group II); electrons injected during the later period (Group III). Below we discuss the properties of these three electron groups as deduced from our PIC simulations. The basic physics underlying the differences between Groups II and III are discussed in Section~\ref{sec:analytics}.

The prominent feature of Group I electrons is the huge injection spike at $x_0 \approx 0.92$mm ($x_0 \approx 100\lambda_L$), where plasma density profile transitions from a linear ramp to a plateau~\cite{Feiyu,Weikum}. This injection is consistently observed for a wide range of peak laser pulse powers $P_L$ and for all polarizations (i.e. all values of $\epsilon$). Group I electrons are injected and gain significant energy ($\gamma \sim 300$) before the start of plasma bubble undulations around $ct \approx 400 \lambda_L$. Therefore, plasma bubble undulations do not have any significant effect on either injection or subsequent dynamics of the Group I electrons. While their energy and charge reach $\gamma_I \sim 500$ and $Q_I \sim 1.3 {\rm nC}$, respectively, their transverse momenta remain moderate ($p_{\perp} \sim 2mc$). 
The energy spread of Group I electrons is fairly large due to beam loading.

Injection of Group II electrons takes place from the plasma region between the two vertical dashed lines shown in Figs.~\ref{fig:injection}(a-c). The primary injection mechanism during this period is the rapid expansion of the bubble's longitudinal size~\cite{kalmykov_prl,austin_ppcf}. This can be observed from the CP laser case shown in Fig.~\ref{fig:injection}(c), where the injected electrons (blue dots) originate symmetrically in their original $z_0$-location.

Despite the relatively long duration ($\Delta \xi/c\approx 370 \mathrm{fs}$: see Figs.~ \ref{fig:injection}(d)-(f)) of the entire bunch train -- or a single bunch in the CP laser case -- Group II electrons collapse into a monoenergetic bunch [Fig \ref{fig:spectra} (c)] at the final propagation distance $x_{\rm fin} \approx 6.4 {\rm mm}$ due to phase space rotation~\cite{kalmykov_prl, austin_ppcf}. At the same time, injection dynamics becomes dependent on the laser polarization after the onset of laser pulse steepening and plasma bubble undulations: see Figs.~\ref{fig:injection}(a)-(c).

The effect of plasma bubble undulations on electron injections into a simultaneously expanding plasma bubble is most transparently illustrated by the LP laser case shown in Fig.~\ref{fig:injection}(a), where highly asymmetric electron injection in $z_0$ can be observed for any given initial electron position $x_0$, i.e. the $(x_0,z_0)$ coordinates of the injected electrons are highly correlated. While electrons injected from a bubble driven by the LP pulse are injected in short bursts from alternating locations with $z_0\approx \pm r_{\rm b}$, those injected into a CP laser driven bubble originate from a spiral-shaped initial positions. Injection rates plotted as red lines in Figs.~\ref{fig:injection}(a)-(c) show that as the laser polarization changes from LP to EP to CP, the rate of electron injections versus the injection position $x_0$ transitions from short periodic bursts to a near-constant value.

Polarization-defined difference between Group II electrons is also manifested in the amount of charge injected by the LP and CP pulses: $Q_{II}^{\rm LP} \approx 3.3 {\rm nC}$ for the former and $Q_{II}^{\rm CP} \approx 3.7 {\rm nC}$ for the latter. The dips in the injection rate (Fig.~\ref{fig:sim}(e)) shows that the high-amplitude bubble undulations produced by the steepened LP pulse destructively interferes with electron injection produced by the elongation of the plasma bubble. $Q_{II}^{\rm LP} < Q_{II}^{\rm CP}$ suggests that less destructive interference appears to be happening for the CP pulse despite the emergence of helical bubble undulation after the CP pulse steepens, as illustrated by Fig.~\ref{fig:sim}(e). A simple mathematical model of such interference is proposed in Sec.~\ref{sec:analytics}.

The most stark difference in the electron injection dynamics produced by the LP, EP, or CP laser pulses can be observed for Group III electrons. Injected during the later (post-steepening) stage of laser pulse propagation, Group III electrons form high-charge high-current electron bunches in the $\xi < 30 \lambda_L$ region as shown in Fig.~\ref{fig:injection}(d,e) for the LP and EP cases. Such current bunching with periodicity $\Delta \xi \sim \lambda_L$ directly reflects multiple periodic electron injections from the $x_0 > 300 \lambda_L$ region of the plasma, with the injection periodicity $T_{\rm CEP}\approx 100\lambda_L/c$ observed in Figs.~\ref{fig:injection}(a,b).

This numerically observed injection periodicity was found to be close to the theoretical estimate of $T_{\rm inj}\approx T_{\rm CEP}/2$. The injected electrons are rapidly accelerated inside the bubble after the injection, and their relativistic factor rapidly increases from $\gamma \approx \gamma_{\rm bb}$ at the injection time to $\gamma \gg \gamma_{\rm bb}$ where $\gamma_{\rm bb} \sim 5$ is the relativistic factor corresponding to the plasma bubble's rear velocity $v_{\rm bb} \equiv \sqrt{\gamma_{\rm bb}^2 - 1}/\gamma_{\rm bb}$. Such periodic injections are expected to translate into a series of current spikes separated by $\Delta\xi \approx cT_{\rm CEP}/4\gamma_{\rm bb}^2 \sim \lambda_L$. This is roughly in agreement with the bunch modulation period of approximately $\Delta \xi /c =1.3\lambda_L/c \approx 40 {\rm fs}$ as observed in Figs.~\ref{fig:injection}(a,b).

The injected total charges of Group III electrons for the LP and CP laser pulses are $Q_{III}^{\rm LP} \approx 7 {\rm nC}$ and $Q_{III}^{\rm CP} \approx 5.6 {\rm nC}$, respectively. In the case of LP laser pulses, each of the current spikes carries approximately $\delta Q_{III}^{\rm LP} \approx 1 {\rm nC}$ of charge and has the duration of the order of $\delta \xi \sim 16 {\rm fs}$. In the case of a CP laser pulse, the electron current is much more uniformly distributed, with the peak current never exceeding $I_{III}^{\rm CP} \approx 45 {\rm kA}$. On the other hand, the peak currents for the LP pulse approach $I_{III}^{\rm LP} \approx 75 {\rm kA}$. The mathematical description of the interplay between electron injections due to plasma bubble undulations and expansion is presented in Section~\ref{sec:analytics}, where we uncover the differences between Group II and III electrons.

Note that even for electron injections driven by a CP laser pulse, there is an injection rate dip at $x_0 \approx 4.8$mm($x_0 \approx 520\lambda_L$: see Fig.~\ref{fig:injection}(c)). This occurs because the peak accelerating gradient inside the plasma bubble is reduced via beam-loading of the plasma wake~\cite{beamloading} by the large earlier-injected electron charge, thereby suppressing further electron injections.

We further remark that the abrupt beam loadings of the plasma wake by short bursts of injection in the case of the LP pulse [Fig \ref{fig:injection} (a)] leads to a slightly wider energy spectrum peak width than for the CP pulse: $\Delta E^{\rm LP}/E_{\rm mono} \approx 4\%$ versus $\Delta E^{\rm CP}/E_{\rm mono} \approx 2\%$ FWHM as can be observed from Fig.~\ref{fig:spectra}(c). Furthermore, different transverse wakefields and injection processes for the LP and CP laser pulses lead to substantially different electron distributions in phase space shown in Fig.~\ref{fig:spectra} (a-b). This, in turn, affects the emitted X-ray as will be discussed later in Sec.~\ref{section:discussions}.

\section{Expanding Phase-dependent Undulating Bubble (EPUB) injection mechanism }\label{sec:analytics}

To interpret this electron injection into an evolving plasma cavity, we use a simplified model of a  positively-charged (devoid of electrons) spherical plasma bubble~\cite{kalmykov_prl,austin_ppcf,kost_injection, kostyukov_pop04}. The bubble has radius $R(t)=R_0(1+\varepsilon t)$ with initial radius $R_0$ expanding with rate $\varepsilon$ propagating with uniform velocity $v_{\rm b}$. A Hamiltonian describing plasma electrons' interaction with the bubble can be written as $H(\boldsymbol{\rho},t) = \sqrt{1+(\mathbf{P}+\mathbf{A(t)})^2} - v_b P_x - \phi(t) $, where $\boldsymbol{\rho} = (\xi,y,\tilde{z})$, $\xi=x-v_b t$, $z_{\rm osc}(t)$ is the transverse coordinate of the undulating bubble center, $\tilde{z}=z-z_{\rm osc}(t)$ is the electron z-coordinate from the undulating bubble center, $\mathbf{P}$ is the canonical momentum, and $\mathbf{A(t)}$ ($\phi(t)$) are the vector (scalar) potentials. Time, length, potential, and electron momentum are normalized to $\omega_p^{-1}$, $k_p^{-1} = c/\omega_p$, $m_e c^2/|e|$, and $m_e c$, respectively, where $\omega_p=\sqrt{4\pi e^2 n_p/m_e}$ is the electron plasma frequency and $n_p$ is the plasma density.

We use the $A_x(t)= -\phi(t) = \Phi(t)/2$  gauge, and assume that $\Phi(t)=(\rho(t)^2-R(t)^2)/4$ inside and $\Phi(t)=0$ outside the bubble. Transverse plasma bubble undulations $z_{\rm osc}(t) \equiv z_{\rm u}\cos(\omega_{\rm CEP} t + \phi_{\rm CEP})$  and bubble expansion, $R(t)$, introduces time dependence of the Hamiltonian. Here $\omega_{\rm CEP}\equiv 2\pi/T_{\rm CEP}$ is the CEP slip rate, $z_{u}$ is the maximum bubble oscillation amplitude, and $\phi_{\rm CEP}\equiv \phi_{\rm CEP}(t(x_0),x_0)$ is the initial CEP evaluated at the time $t(x_0)$ corresponding to electron's entrance into the bubble at $x=x_0$.

To simplify the discussion, we consider the electron motion in the x-z plane. From the Hamiltonian, the  equations of motion (see \ref{Hamiltonian}) and the following Hamiltonian time-dependence can be derived:

\begin{eqnarray}
\label{eq:eqn5}
&\frac{d H}{dt} =-\frac{1+v_x}{4}\left[\tilde{z}(t)\dot{z}_{\rm osc}(t)+R(t)\dot{R}(t)\right].
\end{eqnarray}

In the following section, this Hamiltonian evolution will be used to determine if the electrons will be injected or not.
\subsection{Analytic estimates of electron injection conditions using Hamiltonian model}\label{subsec:Hamiltonian_injection}
Under specific conditions, electrons get injected into the bubble and are accelerated to ultra relativistic energies 
~\cite{kalmykov_prl, austin_ppcf, kost_injection}. An electron can be trapped when the condition $H<0$ is fulfilled ~\cite{kalmykov_prl}; under this condition, electrons cannot escape the bubble, even when they overtake the  bubble. There is another population of electrons, the injected electrons, which can gain similar peak energy in the bubble as the trapped electrons but can escape the bubble after it reaches the front of the bubble ~\cite{austin_ppcf}. In this paper, we  present an estimate of this injection condition, providing a slightly relaxed condition for electrons to enter and gain relativistic energy.

 Injection condition for a moderately relativistic bubble ($R \sim \gamma_{\rm b}$) where $\gamma_{\rm b}$ is the bubble's relativistic factor, was derived using simplified equations of motion~\cite{kost_injection}. Electrons can catch up with the bubble if they can reflect off the bubble's rear wall at least once. This results in electron spending longer time in the bubble, and can determine if the electrons will be injected or not. This reflection and subsequent injection was shown to be true when $\sqrt{2}\gamma_{\rm b}<R$ .

In an expanding bubble, this condition can be relaxed because the Hamiltonian of the electron is altered~\cite{kalmykov_prl}.  For a ultra-relativistic electron ($p_x\gg 1$) interacting with an ultra-relativistic ($\gamma_{\rm b} \gg 1$) non-evolving bubble, maximum excursion of electron from bubble's center axis is given by  $r_{\rm m}\approx 4H+R^2-2p_x/\gamma_{\rm b}^2-2/p_x$, with $H=1$ ~\cite{kost_injection}. Assuming small expansion rate ($\varepsilon \ll 1$) of a bubble with a time-dependent radius $R(t)$ expanding according to $R(t)=R_0 \left( 1 + \varepsilon t \right)$, the maximum momentum gained by the electron is almost identical as for the static bubble ($\varepsilon=0$). Therefore, $r_{\rm m}$ can be simply modified by using a modified (instantaneous) $H(t)$.

Numerical solutions of normalized equations of motions Eqns (\ref{eq:eqn6}-\ref{eq:eqn9}, See Appendix B) shows that an initially quiescent ($P_{x0}=P_{z0}=0$) electron entering a non-evolving bubble from the top edge ($dP_{z0}/ds=-1/4, dP_{x0}=0, X_0=0, Z_0=R$), gains maximum longitudinal momentum $p_x\approx 1.1 R^2$ before leaving bubble~\cite{kost_injection}.  Combining this with $r_{\rm m}<R$, a modified injection condition can be found:
\begin{equation}
    \gamma_{\rm b}/R<1.1/\sqrt{2H}.
\end{equation}
Or alternatively,
\begin{equation}
    H<H_{\rm thresh}\approx 0.6R^2/\gamma_{\rm b}^2.
\end{equation}
For an ultra-relativistic bubble where $\gamma_{\rm b}\gg R$, this will only hold when $H\approx 0$, in accordance with the previous result ~\cite{austin_ppcf}, and electron trapping~\cite{kalmykov_prl} becomes necessary. For an initially quiescent electron $H_0=1$ to get injected, the change of Hamiltonian, $\Delta H = H-1 < \Delta H_{\rm thresh}= H_{\rm thresh}-1$ needs to hold.

When an electron interacts with an undulating bubble, its Hamiltonian can further increase or decrease, since the term $\dot{z}_{\rm osc}$  in $dH/dt$ will change sign according to period $T_{CEP}~\cite{Jihoon}$. This change of electron Hamiltonian can trigger or suppress electron injection at sub-optimal (optimal) undulation phases, $\phi_{CEP}=\pi/2$ ($-\pi/2$) for $\tilde{z}=-R$ and $\phi_{CEP}=-\pi/2$ ($\pi/2$) for $\tilde{z}=R$, by increasing (decreasing) the Hamiltonian above the injection threshold. The combined effect of expansion and undulation can result in periodic electron injections from the background plasma, which in turn modifies the injected bunch current profile.

\begin{figure}[h]
\centering
    \includegraphics[width=0.8\textwidth]{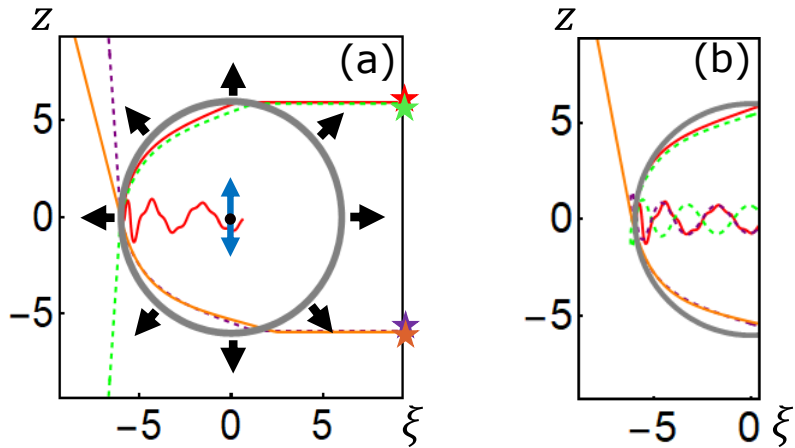}
\caption{Electron interaction with an evolving bubble: single-particle simulations. (a,b) Trajectories of background electrons interacting with slowly (a) and rapidly (b) expanding plasma bubbles: with (solid lines) and without (dashed lines) simultaneous bubble undulations. Bubble expansion/undulation is marked by black/blue arrows. All electrons start at $(\xi_0=8/k_p,z_0 = \pm 6/k_p)$ initial positions (stars). Bubble parameters: $R_0=6/k_p$, $\xi_0=8$(stars), $\varepsilon=0.001 $ (a), $\varepsilon = 0.003$ (b), $z_{\rm u}=1.5/k_p$, $T_{\rm CEP}=50/\omega_p,  \phi_{CEP}=\pi/2$.  }\label{fig:trajectories}
\end{figure}

To illustrate the modified injection condition and the effect of bubble expansion and undulation, we solve equations of motion (Eqns \ref{eq:eqn1}-\ref{eq:eqn4}, see Appendix B) for four initially quiescent electrons entering the bubble at $y=0, \xi=x-v_{\rm b} t =0, z=\pm R$ [Fig. \ref{fig:trajectories}]. The bubble oscillates with a period $T_{\rm CEP}=50$, and also expands at different rate until $t_{\rm exp}=20$. For a slow-Expanding Bubble(EB) that does not undergo undulations (dashed lines in Fig \ref{fig:trajectories}(a): $z_{\rm u}=0$), the expansion rate $\varepsilon=0.001$ is insufficient to cause injection. For a finite undulation amplitude (solid lines in Fig.~\ref{fig:trajectories}(a): $z_{\rm u}=1.5/k_p$), injection can be enabled, but only for certain undulation phases $\phi_{\rm CEP}$ or impact parameters $z_0$. For example, for a given undulation phase, injection into an EPUB is enabled for the electron with $z_0 = R_0$ (red solid line), but not for the electron with $z_0 = - R_0$ (orange solid line). This example of a slowly-expanding undulating plasma bubble emulates the injection mechanism of Group III electrons described earlier in Section~\ref{sec:pic}.

For a fast-EB exemplified by Fig.~\ref{fig:trajectories}(b), where the expansion rate is chosen as $\varepsilon=0.003$, bubble undulations can play the opposite role of suppressing electron injections. For such expansion rate, electrons are injected (dashed lines) when the bubble is not undulating. Similar to the slow-EB case, the electron with ``optimal" impact parameter $z_0 = R_0$ (red solid line) is injected, but the one with the``sub-optimal" impact parameter $z_0 = - R_0$ (orange solid line) is prevented from injection by a finite-amplitude plasma bubble undulation. This example of a rapidly-expanding undulating plasma bubble emulates the injection suppression mechanism of Group II electrons described in Section~\ref{sec:pic}.

\begin{figure}[h]
    \includegraphics[width=\textwidth]{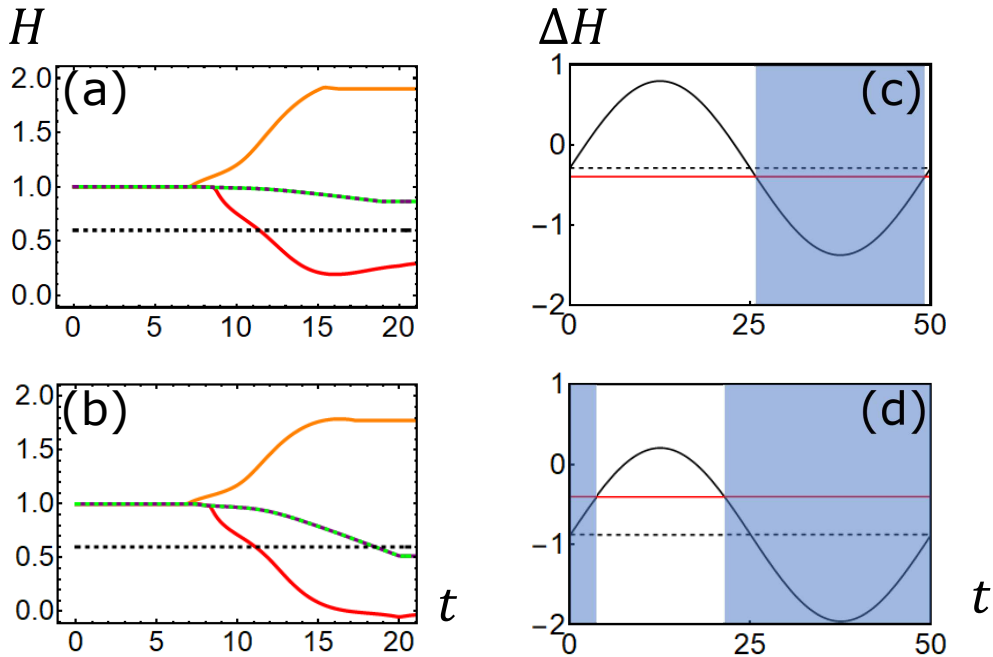}
\caption{(a,b) Time-evolution of the electron Hamiltonian $H(t)$ for slow-EB (a) and fast-EB (b): no undulations ($z_{\rm u}=0$). Initial conditions: same as in Fig. \ref{fig:trajectories}. Black dashed lines: injection threshold Hamiltonian $H_{\rm thresh}$. (c-d) Time-evolution of the electron Hamiltonian increment $\Delta H(t)$ for undulating ($z_{\rm u}=1.5/k_p$) slow-EB (c) and fast-EB (d). $\Delta H^{(1)}$ (Eqn. \ref{eqn:H_estimate}, Black Solid), $\Delta H_{\rm exp}$ (Black Dashed) , $\Delta H_{\rm thresh}=H_{\rm thresh}-1$ (Red). Injection is enabled for those $0<t< T_{\rm CEP}$ inside blue-shaded regions.  Bubble parameters: same as in Fig.~\ref{fig:trajectories}}.\label{fig:Hamiltonian}
\end{figure}

This difference can be understood more quantitatively by tracking the electron Hamiltonian. For slow-EB, Electron Hamiltonian does not decrease below $H_{\rm thresh}$ without undulation, regardless of initial impact parameters [Fig \ref{fig:Hamiltonian} (a), Green Dashed, Purple Dashed]. Only the electron experiencing optimal phase undulation and expansion [Fig \ref{fig:Hamiltonian} (a), Red] can have $H<H_{\rm thresh}$ and get injected, in accordance with particle trajectory shown in Fig \ref{fig:trajectories} (a).

For fast-EB, expansion is fast enough so the Hamiltonian decreases below $H_{\rm thresh}$ even without undulation [Fig \ref{fig:Hamiltonian} (b), Green Dashed, Purple Dashed]. For this Scenario, electron experiencing the sub-optimal phase undulation and expansion is the only one that is not injected [Fig \ref{fig:Hamiltonian} (b), Orange] since its Hamiltonian increases above $H_{\rm thresh}$.

We have discussed electron injection from an evolving bubble for optimal and sub-optimal phases. To better understand injection of background particles for all different phases, we give an estimate for final Hamiltonian for optimally positioned electrons($\tilde{z}=\pm R$)~\cite{kalmykov_prl,kost_injection} entering the bubble as the bubble propagates using estimates from Eqns (\ref{eq:eqn6}-\ref{eq:eqn9}). We compute Hamiltonian of electrons located on the sinusoidal trajectory defined by  $x_0=v_{\rm b} t, y_0=0, z_0=\pm R+z_{\rm osc}(t)$. Electrons located on this trajectory graze the bubble boundary, entering the bubble at its edge.  Change in Hamiltonian of electrons entering bubble at different time can be estimated by integrating

\begin{equation}
    \Delta H = \int dt \left[ \dot{p_z}(t)\dot{z}_{\rm osc}(t) -\frac{1+v_x}{4}R(t)\dot{R(t)}\right],
\end{equation}
where the integral is calculated along the  electron trajectory.

To lowest order in bubble oscillation amplitude and expansion rate, we can use the quantities from a non-evolving bubble($z_{\rm u}=0, \varepsilon=0$) to estimate change in Hamiltonian. Assuming passage time of electron through the bubble, $T_{\rm pass}$, is much smaller than oscillation period, $T_{\rm CEP}$, $z_{\rm osc}\approx - z_{\rm u}\omega_{\rm CEP} \sin(\omega_{\rm CEP}(t_{\rm enter}))$, with $t_{\rm enter}$ the time electron enters the bubble. Furthermore, if the bubble radius varies slowly ($\varepsilon\ll1$), $R(t)\dot{R}(t)\approx R^2\varepsilon$. This simplifies the integral to $\Delta H\approx -\left[\mp z_{\rm osc}\omega_{\rm CEP}\sin(\phi_{\rm CEP})\Delta p_z +(t_{\rm exit}+\Delta x)(R^2\varepsilon)/4\right]$ where $\Delta p_z$ is the exit transverse momentum of the electron, $t_{\rm exit}$ is the time at which electron exits the bubble, and $\Delta x$ is the longitudinal distance electron travels during interaction with the bubble.

To estimate $\Delta x+t_{\rm exit}$ and $\Delta p_z$, we solve Eqns (\ref{eq:eqn6} - \ref{eq:eqn9}), obtaining $\Delta x+t_{\rm exit}=5.4 R, \Delta p_z = \mp 0.16R^2$ when the electron reaches is maximum excursion from axis near the back at $r=r_{\rm m}$. Using this result, the Hamiltonian increment for an electron entering the bubble from the edge ($\tilde{z_0}=z_0-z_{\rm osc}(t)=\pm R$) is given by
\begin{equation}
    \Delta H^{(1)}= -1.35\varepsilon R_0^3 \; \pm0.16 z_{\rm u}\omega_{\rm CEP} R_0^2 \times \sin(\omega_{\rm CEP}t_{\rm enter}),
    \label{eqn:H_estimate}
\end{equation}
where the two signs correspond to the two impact parameters $\tilde{z_0}=\pm R$. Equation (\ref{eqn:H_estimate}) contains two terms with distinct behaviors: the bubble expansion contribution $\Delta H_{\rm exp}=-1.35\varepsilon R_0^3 $ and the bubble undulation contribution $\Delta H_{\rm osc}= \pm 0.16 z_{\rm osc}\omega_{\rm CEP} \sin(\omega_{\rm CEP}t_{\rm enter})R_0^2$. The former depends on the expansion rate of the bubble and does not depend on the time $t_{\rm enter}$ of the electron encounter with the bubble. On the contrary, the latter terms depends on the bubble undulation amplitude, and is periodic in $t_{\rm enter}$ with a period $T_{\rm CEP}$.

Therefore, bubble undulations can affect electron injection in two ways. {\it Constructive contribution of bubble oscillations} to electrons injection occurs when its expansion rate $\varepsilon$ is not sufficiently large to cause injection: $1+\Delta H_{\rm exp}>H_{\rm thresh}$. In that case (see black dashed and red solid lines in Fig.~\ref{fig:Hamiltonian}(c)), finite $\Delta H_{\rm osc}$ can further reduce $H$ and cause electron injection. If the oscillation amplitude of $\Delta H_{\rm osc}$ is large enough, bubble oscillations can reduce the Hamiltonian below $H_{\rm thresh}$ for some injection times $t_{\rm enter}$. As indicated by blue shading in Fig.~\ref{fig:Hamiltonian}(c), this would result in periodic electron injections during the time periods when $\Delta H^{(1)}$ (solid black line) drops below the injection threshold $H_{\rm thresh}$ (solid red line). In principle, the intervals of electron injection can be as short as possible because the time intervals during which $H<H_{\rm thresh}$ can be arbitrarily short. Therefore, electrons injected owing to constructive contribution of bubble oscillation correspond to Group III electrons previously described in Section \ref{sec:pic}.

{\it Destructive contribution of bubble oscillations} to electrons injection occurs when $\varepsilon$ is large enough to cause injection on its own: $1+\Delta H_{\rm exp} < H_{\rm thresh}$, as indicated by black dashed and red solid lines in Fig.~\ref{fig:Hamiltonian}(d). However, finite $\Delta H_{\rm osc}$ will periodically increase $\Delta H$, thereby suppressing injection for at least some time periods shaded white in Fig.~\ref{fig:Hamiltonian}(d). We note that the injection periods (blue-shaded region in Fig.~\ref{fig:Hamiltonian}(d)) cannot become arbitrarily short because the time periods during which injection is suppressed cannot be longer than $T_{\rm CEP}/2$. Electrons injected owing to destructive contribution of bubble oscillations correspond to Group II electrons.

The two injection scenarios describing the emergence of Group II and III electrons predict that the injection process is periodic in space along the propagation direction, with a spatial period $L_{\rm CEP} \approx c T_{\rm CEP}$. However, only the first scenario enables electron injection over a propagation distance $L_{\rm inj} < L_{\rm CEP}/2$ through the plasma. This finding illuminates the reason for the short durations $\Delta \xi$ of the beamlets comprising Group III electrons: because the injected electrons (moving with relativistic speed $\approx c$) are slowly slipping with respect to the back of the plasma bubble (moving with the speed $\approx v_{\rm bb}$), plasma electron injected over a distance $L_{\rm inj}$ are compressed into ultra-short bunches with $\Delta \xi \approx L_{\rm inj}/2\gamma_{\rm bb}^2$, where $\gamma_{\rm bb}\approx 1/\sqrt{1-v_{\rm bb}^2/c^2}$. Not surprisingly, ultra-short bunches with $\Delta \xi \approx L_{\rm inj}/2\gamma_{\rm bb}^2$ were observed only for Group III electrons in our PIC simulations, as shown in Fig.~\ref{fig:injection}(d).

Up to now, we have only considered $\Delta H$ in the case of a plasma bubble undulating in one direction. Plasma bubble whose center moves along a helix can be also generated by a circularly polarized laser pulse. Under this circumstance, the bubble centroid would always have a finite undulation speed, although the direction of its transverse motion would be continuously changing. The helical nature of bubble undulations has a direct impact on electron injection into the bubble because at any point in time, there are some directions from which injection is either suppressed or enhanced. The effect of helical undulations of the bubble centroid on the temporal structure of the accelerated beam is investigated below in Section~\ref{subsec:swarm}. We note that it is also possible to use elliptically polarized laser to drive the bubble, which can generate a bubble whose center follows a helix elongated in a specific transverse direction.

\subsection{Polarization-dependent current profiles: particle swarm simulations}\label{subsec:swarm}

The analytic calculations presented above demonstrate the feasibility of controlling electron injection via the combination of plasma bubble expansion and undulation. For analytic tractability, the calculations in Sec.~\ref{subsec:Hamiltonian_injection} are limited to those electrons that are most likely to be injected, i.e. having an impact parameter $r_0 \equiv \sqrt{z_0^2 + y_0^2} = R$ matched to the unperturbed bubble radius $R$. In fact, background electrons with different impact parameters interact with an evolving bubble; some of these electrons get injected into it. Therefore, in order to interpret the results of our PIC simulations presented in Sec.~\ref{sec:pic}, it is necessary to simulate a large-volume ``swarm" of background plasma electrons interacting with a plasma bubble. This was done by seeding test particles into a three dimensional volume spanning a wide range of initial conditions ($R<x_0<80, -10.5< y_0, z_0 <10.5$) and launching three different evolving spherical potentials that can capture and accelerate particles: (1) an expanding bubble (EB), (2) an expanding and helically undulating bubble (EHUB), and (3) an expanding and linearly undulating (ELUB).

In the EHUB case, the center of the bubble moves transversely according to $z_{\rm osc}(t) \equiv z_{\rm u}\cos(\omega_{\rm CEP} t + \phi_{\rm CEP}), y_{\rm osc}(t) \equiv z_{\rm u}\sin(\omega_{\rm CEP} t + \phi_{\rm CEP})$. There is no transverse bubble undulation ($z_{\rm u} = 0$) in the EB case. The 3D particle equations of motion are similar to Eqns.~(\ref{eq:eqn1}-\ref{eq:eqn4}) derived for a particle traveling in the $x-z$ plane.

\begin{figure}[h]
    \includegraphics[width=\textwidth]{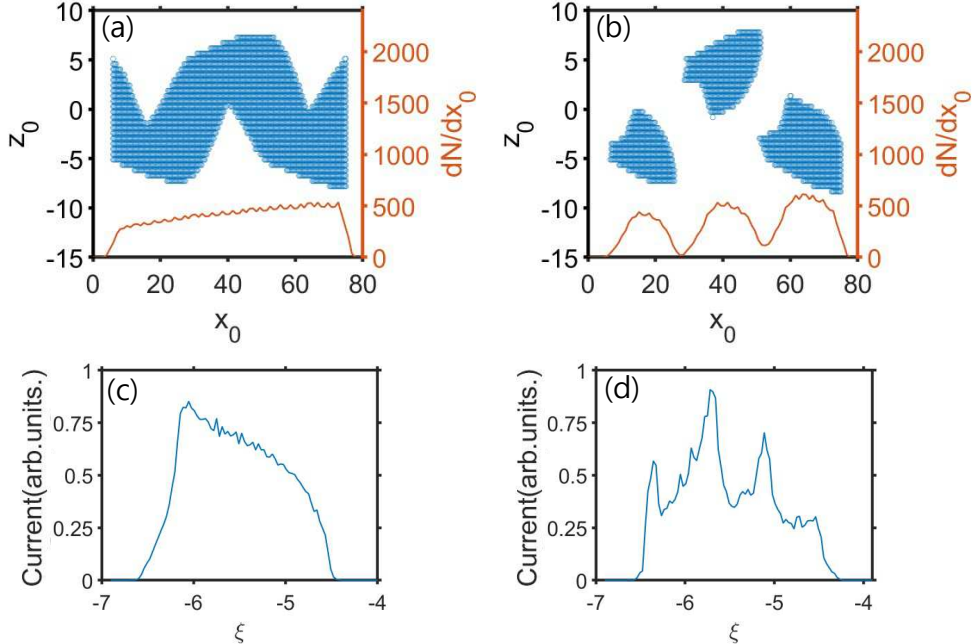}
\caption{Injection and phase space modeling using particle swarm simulations. (a-b) Injected particles plotted in  $x_0$-$z_0$ plane(blue dots) and injection rate (red line) and  (c-d) current profile of injected particles at $t=150$ for EHUB [(a),(c)] and ELUB [(b),(d)]. Normalized bubble parameters:  $R_0=6, \gamma_{\rm b}=6, \varepsilon=0.002, z_{\rm u}=1.5$ }\label{fig:swarm}
\end{figure}

As can be seen from the injection rate~[Fig \ref{fig:swarm} (a)-(b)], only the ELUB injection rate is periodically modulated. While the injection rate of an EHUB does not exhibit modulation, the transverse location from which they originate from does show periodic modulation [Fig \ref{fig:swarm} (a)]. Injected electron distribution in Fig \ref{fig:swarm} (a)-(b) and injection rate in Fig \ref{fig:swarm} (d) are periodic. Their initial location projected onto x-z plane has an approximate periodicity of $cT_{\rm CEP}\approx 50$, in agreement with the time dependence of  $\Delta H^{(1)}$. We note that injection rate for ELUB has periodicity $cT_{\rm CEP}/2$, since the injection process happens twice, at $\tilde{z}\approx \pm R$, for each undulation period.

After the electrons are injected into the bubble, they quickly gain relativistic energy from the accelerating field and move at ultra relativistic velocity. Because the bubble phase velocity is slower than that of the injected electrons, electrons will advance through the bubble after  acceleration to ultra relativistic energy. This slippage of the back of the bubble from the injected bunches determines the longitudinal structure of the injected bunch in the case of a linearly undulating bubble. One can estimate the periodicity of the bunch modulation via converting injection periodicity to that in the moving-frame $\xi=x-v_{\rm b} t$. The rear of the bubble moves at $v_{\rm bb} = v_{\rm b} - R\varepsilon$ slower than bubble velocity because of bubble expansion. While the CEP phase slips one cycle, the back of the bubble slips away from the ultra-relativistic particles by distance $\Delta\xi=(c-v_{\rm bb})T_{\rm CEP}$. Because there are two injection per one oscillation, injected bunch forms a structure with longitudinal modulation  $\Delta\xi=(c-v_{\rm bb})T_{CEP}/2\approx T_{\rm CEP}/4\gamma_{\rm bb}^2$~[Figure \ref{fig:swarm}(d)]. We note that this is the formula that was cited for bunch modulation periodicity in Section \ref{sec:pic}.

\subsection{Effect of bubble undulations on the betatron radiation emitted by injected/accelerated electrons}\label{subsec:Xrays}

\begin{figure}[h]
    \includegraphics[width=\textwidth]{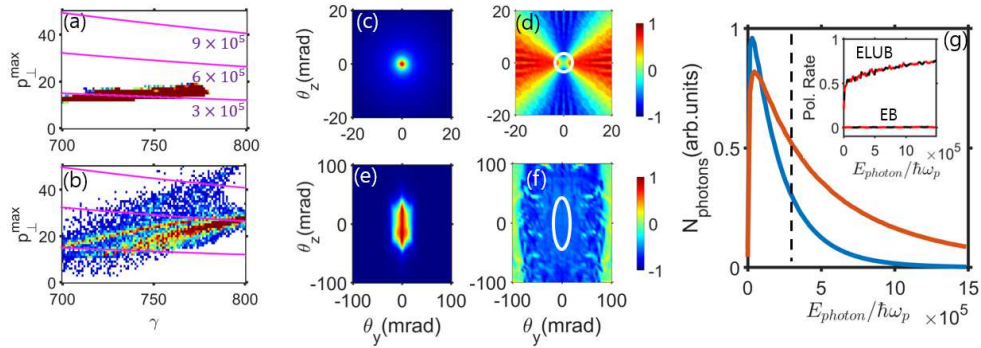}
\caption{Properties of betatron X-ray radiation calculated from particle swarm simulations. (a-b) Electron phase space at $t=400$ propagation for EB (a) and ELUB (b). Pink lines: critical X-ray frequencies $E_{\rm crit}/\hbar \omega_p$ calculated from Eq.~(\ref{eq:critical_freq}). (c-f) Angular distribution of the X-ray intensity (c,e) and the first Stokes parameter $S_1/S_0$ (d,f) calculated from electron trajectories inside EB (c-d) and ELUB (e-f). White lines in (d,f): half-max intensity contour $I=I_{\rm max}/2$. All X-ray energies are included in (c,e). Fixed X-ray energy $E_{\rm X-ray}/\hbar \omega_p = 3\times 10^5$ (dashed line in (g)) is used in (d,f).  (g) Total X-ray intensities from EB (blue) and ELUB (red). Inset: on-axis degrees of overall polarization $P=\sqrt{ S_1^2+S_2^2+S_3^2}/S_0$ (red dashed line) and linear polarization $|S_1/S_0|$ (black line).} \label{fig:test_radiation}
\end{figure}

Laser-wakefield generated electrons can emit collimated, high-brightness X-rays via betatron radiation~\cite{Felicie_radiation}. Because the injected electrons form beams with femtosecond-scale durations, the resulting pulses of betatron radiation have a similarly ultrashort temporal format and, potentially, tunable polarization imparted by that of the laser pulse~\cite{keV_Xray,dopp_lsa17,svensson_nphys21}. Such X-rays have been used to image various targets with fine details. Some of the recent examples include irregular eutectics in the aluminum-silicon (Al-Si) system~\cite{Amina}, as well as various biological samples~\cite{bee}. Below we demonstrate that the EPUB injection mechanism provides a new approach to controlling X-ray polarization, intensity, and angular distribution.

Our particle swarm simulation shows that electrons injected by an expanding and undulating bubble gain more transverse momentum than those injected into a merely expanding bubble. We plot the electron distribution in the $(\gamma,p_\perp^{\rm max})$ space at propagation distance of $t=400$ for expanding~[Fig \ref{fig:test_radiation} (a)] and expanding and linearly undulating~[Fig \ref{fig:test_radiation}(b)] bubbles. Here, $p_\perp^{\rm max} = \sqrt{2\gamma\epsilon_{\perp}}$ denotes the maximum possible transverse momentum derived from the transverse energy $\epsilon_\perp = p_\perp^2/2\gamma + r^2/4$ ~\cite{xi_radiation}.

We observe from Fig.~\ref{fig:test_radiation}(a) that the electrons injected into a merely expanding bubble do not spread out in the $(\gamma,p_\perp^{\rm max})$ phase space, forming a line-like feature. On the contrary, those injected by an ELUB significantly spread out in phase space, as can be observed from Fig.~\ref{fig:test_radiation}(b). This difference has direct consequences for the radiated X-ray spectra because the frequency range of the X-ray photon energy emitted by the electrons is determined by the critical frequency $\omega_{\rm crit}$ that can be estimated as~\cite{xi_radiation}
\begin{equation}\label{eq:critical_freq}
    \omega_{\rm crit} \approx \frac{3\omega_p}{\sqrt{8}} \gamma^{3/2}p_\perp.
\end{equation}
Therefore, an undulating bubble could potentially yield higher-energy X-rays because of the larger values of $p_{\perp}$. Also, a wider angular spread is predicted for X-ray generated from undulating bubble because synchrotron-like radiation has an opening angle of $\theta_{\rm emission} \approx p_\perp/\gamma$ ~\cite{Jackson}.

To compare X-ray emission from the non-undulating and undulating bubble, we have calculated the betatron radiation from a swarm of relativistic test particle trajectories according to the standard expression~\cite{Jackson,VDSR}:
\begin{equation}
    \frac{d^2I}{d\omega d\Omega}=\frac{e^2\omega^2}{4\pi^2c} \left|\int_{-\infty}^{\infty} \mathbf{n \times\left( n\times\boldsymbol{\beta}\right) e^{i\omega(t- \mathbf{n\cdot r(t)}/c)}}\right|^2.
\end{equation}
The integration was carried out using an in-house code \href{https://github.com/tianhongg/SIRC}{Simple Incoherent Radiation Calculation (SIRC)}~\cite{SIRC}. We compute the Stokes parameters according to the formula $I=S_0=E_y\cdot E_y^*+E_z\cdot E_z^*,
S_1=E_y\cdot E_y^*-E_z\cdot E_z^*,
S_2=E_y\cdot E_z^*+E_z\cdot E_y^*,
S_3=-i(E_y\cdot E_z^*-E_z\cdot E_y^*)$. The polarization degree is expressed as $P=\frac{\sqrt{S_1^2+S_2^2+S_3^2}}{S_0^2}$~\cite{Jie_Feng}, and the fraction of vertical and horizontal polarization may be characterized by $P_1=S_1/S_0$.

Because electrons are symmetrically injected into an EB, they undergo betatron oscillations in the $y-z$ plane without any directional preference, and without significant transverse momentum spread, as indicated in Fig.~\ref{fig:test_radiation}(a) by their vertical clustering. Therefore, their betatron (X-ray) emission is symmetric and confined in small angular region with FWHM $\Delta \theta_y = \Delta \theta_z \sim 10$ mrad, as observed in Fig.~\ref{fig:test_radiation}(c). The normalized Stokes parameter $S_1/S_0$ characterizing the degree of linear polarization of the resulting X-rays is plotted in Fig~\ref{fig:test_radiation}(d). We observe that the X-rays  are essentially un-polarized ($S_1/S_0 \approx 0$) near the axis ($\theta_y = \theta_z \approx 0$: inside the white circle), where the radiation intensity is the highest. While direction-dependent linear polarization is observed at larger emission angles, the number of such X-rays is small.

The situation is qualitatively different for the electrons injected into an ELUB undulating along the $z$-direction. These electrons, initially trapped near the axis, subsequently experience transverse kicks in the undulation direction and start executing large betatron oscillations that are predominantly along the $z$-direction. Consequently, the X-rays are emitted with anisotropic angular distribution as shown in Fig.~\ref{fig:test_radiation}(e): $\Delta \theta_y \ll \Delta \theta_z \sim 70$ mrad. Moreover, the on-axis X-rays are strongly polarized in the undulation direction, as can be observed in Fig.~\ref{fig:test_radiation}(f): $S_1/S_0 \approx -0.5$.  We observe that the angular intensity distribution of X-rays plotted in Fig.~\ref{fig:test_radiation} strongly correlates with their polarization properties. In the case of an ELUB, the X-rays are primarily polarized in the z-direction according to Fig.~\ref{fig:test_radiation}(f), and their angular distribution is anisotropic, i.e., also elongated in the same direction according to Fig.~\ref{fig:test_radiation}(e). Likewise, the lack of X-ray polarization in the EB case correlates with their isotropic angular distribution.

As discussed earlier, the spectra of the emitted X-rays extend to higher energies for the linearly-undulating (ELUB) bubble vs the ellongating bubble without undulations (EB), as indicated by the red (for ELUB) vs black (for EB) lines in Fig.~\ref{fig:test_radiation}(g). Also, the on-axis X-rays from the ELUB are more than $50\%$ polarized in the undulation direction, in comparison to almost unpolarized on-axis X-ray from the EB non-undulating bubble: see the inset of Fig.~\ref{fig:test_radiation}(g). Notably, the resulting high degree of linear polarization is achieved without introducing a tilt in the laser pulse front, or using asymmetric laser intensity distribution~\cite{schnell_ncomm16}. Instead, the X-rays are linearly polarized because they are emitted via betatron radiation by trapped/accelerated plasma electrons subjected to the transverse wake $\bf{W}_{\perp} \approx \bf{e}_z \rm{W_z}$ that originates from laser polarization controlled undulations of a plasma bubble.

\section{Potential applications and telltale signs of phase-dependent laser-plasma acceleration}\label{section:discussions}
In this Section, we discuss possible applications of electron injection and acceleration using the EPUB approach. Those include highly efficient generation of high-charge high-energy electron beams and the control of the electron beam pointing by the CEP offset. We will also describe X-ray generation by the electrons produced using differently-polarized laser pulses, as well as the effect of the laser polarization and its CEP offset on the X-ray properties, such as their angular distribution. Unlike Sec.~\ref{subsec:Xrays}, where single-particle simulations of electron injection and acceleration in prescribed fields were used, here we use first-principles PIC simulations to model X-ray generation. Therefore, collective effects such as wake depletion and subsequent transition from LWFA to PWFA regimes, are properly accounted for~\cite{ALLS_PWFA, High_Charge}. Many of the observables described below, such as finite electron pointing angle from the laser axis and angular asymmetry of the emitted betatron radiation, can be used as telltale signs of phase- and polarization-dependent effects in ultra-intense laser plasma interactions~\cite{CEP_observable}. In the rest of this Section, we assume the same laser and plasma parameters as used in Sec.~\ref{sec:pic}, listed in the caption of Fig.~\ref{fig:sim} and Table \ref{table1}. Energy spectra of the accelerated electrons are shown in Fig.~\ref{fig:spectra}(c). For conceptual simplicity, only linear and circular polarizations of the driver laser are considered in this Section.

One of the key desirable metrics of any Laser Plasma Wakefield Accelerator (LPWA) scheme is the high energy conversion efficiency of the laser energy into the energy of accelerated electrons. It has been shown that few-cycle driven lasers can efficiently transfer their energy into the injected electrons~\cite{Papp}. In our simulations presented in Fig.~\ref{fig:spectra}, electrons are injected almost as soon as the pulse enters plasma, injecting extremely high charge ($Q \sim 10 {\rm nC}$) into the wakefield. The large injected charge can efficiently convert the wakefield energy excited by the pulse into electron kinetic energy. Furthermore, the pulse is almost depleted by the time the monoenergetic peak reaches $\approx 200$MeV as shown in Fig.~\ref{fig:spectra}(c). Therefore, the depletion and dephasing lengths are well-matched, and almost no pulse energy is wasted because of the injected electrons entering the decelerating portion of the bubble. This results in high efficiency ($\sim50\%$) of laser energy conversion into electron kinetic energy. Below we discuss how the resulting high-charge ultra-relativistic near-monoenergetic electron bunch can be utilized for producing large amounts of polarization-dependent and CEP-dependent X-rays via betatron radiation.

\subsection{Radiation generation and beam asymmetry}\label{subsec:beam_asymm}

\begin{figure}[h]
    \includegraphics[width=\textwidth]{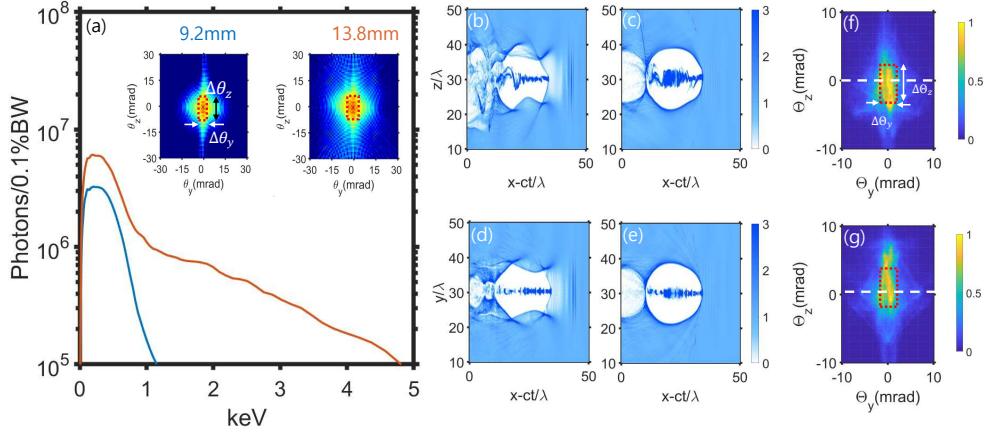}
\caption{CEP- and polarization-dependent asymmetry of the electron pointing directions $(\Theta_y,\Theta_z)$ and their betatron radiation. (a) Spectral brightness of X-rays at $ct_1 = 1,000 \lambda_L$ (blue line) and $ct_2 = 1,500 \lambda_L$ (red line) propagation distances. Insets: angular distributions at $ct_{1,2}$; photon energies $\hbar \omega_{\rm X-ray} < 3 {\rm keV}$. Red boxes at FWHM: $\left( \Delta\theta_y^{(1)} = 4 {\rm mrad}, \Delta \theta_z^{(1)} = 11 {\rm mrad} \right)$ for $ct_1$, $\left( \Delta\theta_y^{(2)} = 6 {\rm mrad}, \Delta \theta_z^{(2)} = 13 {\rm mrad} \right)$ for $ct_2$. (b-e) Normalized plasma density $n/n_0$ in the (b,c) x-z and (d,e) x-y planes for (b,d) $ct_1$ and (c,e) $ct_2$. (f-g) CEP-dependent transverse phase spaces $(\Theta_y,\Theta_z)$ of ultra-relativistic electrons ($\gamma > 380$) at $ct_1$: (f) $\phi_{\rm CEP}=0$ ($\left< \Theta_z \right> = -1 {\rm mrad}$) and (g) $\phi_{\rm CEP}=\pi/2$ ($\left< \Theta_z \right> = 1 {\rm mrad}$). Red boxes at FWHM: $\left( \Delta\Theta_y = 4 {\rm mrad}, \Delta \Theta_z = 8 {\rm mrad} \right)$. White dashed lines: $\Theta_z=0$.}\label{fig:Asymmetry}
\end{figure}

After the formation of a quasi-monoenergetic high-charge bunch,  the laser pulse energy is depleted and can no longer excite a strong wake  at the propagation distance $c t_1 = 1,000\lambda_L = 9.2 {\rm mm}$, as shown in Figs.~\ref{fig:Asymmetry}(b,d). However, the large accelerated charge can now excite its own plasma wake afterwards, resulting in a transition from the LWFA into the PWFA regime~\cite{ALLS_PWFA, High_Charge} as illustrated in Figs.~\ref{fig:Asymmetry} (c,e) for $c t_2 = 1,500\lambda_L = 13.8 {\rm mm}$. A further acceleration of the trapped electrons injected during the later LWFA stages ensues, leading to larger spread but higher peak of electrons energy. According to Eq.(\ref{eq:critical_freq}), higher-energy electrons can produce more energetic X-ray photons via betatron radiation. The transition to PWFA regime results in much larger numbers and energies of the X-rays produced inside the $0 < x < ct_2$ (red line) region than in the $0<x<ct_1$ (blue line) region, as shown in Fig.~\ref{fig:Asymmetry}(a). After laser depletion, bubble undulations essentially stop, as can be observed from a more symmetric plasma bubble shape at $t=t_2$ than at $t=t_1$: compare Figs.~\ref{fig:Asymmetry}(b,c). Nevertheless, several laser polarization- and phase-dependent observables persist even after the laser pulse depletion, i.e. after the LWFA-to-PWFA transition.

First, the electron beam acquires an elongated shape along the direction of the laser polarization, as well as a CEP-depedent average transverse tilt, as shown in Fig.~\ref{fig:Asymmetry}(f-g). Specifically, the FWHM angular electron spread in the $y$- and $z$-directions is given by $(\Delta \Theta_y,\Delta \Theta_z)= \left( 4 {\rm mrad}, 8 {\rm mrad} \right)$ after the propagation distance of $ct_1$ through the plasma. Therefore, the angular spreading of the electron beam is anisotropic, i.e. more extended along the laser polarization direction $z$. This anisotropic angular beam spread is responsible for the similarly anisotropic spread of the emitted X-rays shown in the insets of Fig.~\ref{fig:Asymmetry}(a). Moreover, we note that the average pointing direction of the beam, characterized by $\left< \Theta_z \right> (\phi_{\rm CEP})$, depends on the CEP offset. As shown in Fig.~\ref{fig:Asymmetry}(f,g), the change from $\phi_{\rm CEP} = 0$ to $\phi_{\rm CEP} = \pi$ reverses the overall deflection angle of ultra-relativistic ($\gamma > 380$) electrons from $\left< \Theta_z \right>(0) \approx -1 {\rm mrad}$ to $\left< \Theta_z \right>(\pi) \approx 1 {\rm mrad}$. Even though $\left| \left< \Theta_z \right> \right|$ is smaller than $\Delta \Theta_z$ by nearly an order of magnitude, it still appears to be an experimentally measurable quantity~\cite{seidel_arxiv22,CEP_observable,CEP_beam_pointing}.

Second, the emitted X-rays retain their elongated angular distribution (see the two insets in Fig.~\ref{fig:Asymmetry}(a)) even after propagating the distance of $ct_2$ through the plasma. For example, the ratio of the angular spreads along ($\Delta \theta_z^{(2)} \approx 13 {\rm mrad}$) and perpendicular to ($\Delta \theta_y^{(2)} \approx 6 {\rm mrad}$) the laser polarization direction is $\Delta \theta_z^{(2)} / \Delta \theta_y^{(2)} \sim 2$ at $t=t_2$. While this ratio is somewhat smaller than the one observed at $t=t_1$ ($\Delta \theta_z^{(1)} / \Delta \theta_y^{(1)} \sim 3$), it can be measured and used for estimating the magnitude of the electrons betatron trajectories~\cite{phuoc_prl06,plateau_prl12}.

\begin{figure}[h]
    \includegraphics[width=\textwidth]{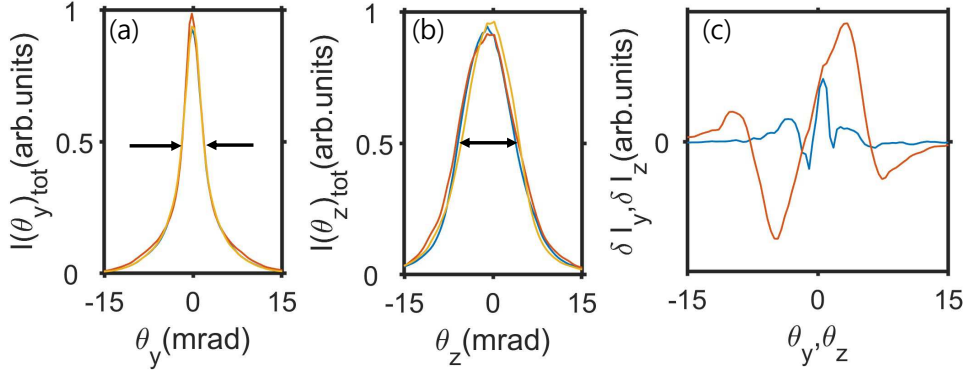}
\caption{CEP-dependent angular X-ray flux distributions for an LP laser.  (a-b) Angle-averaged X-ray fluxes $ \bar{I}_{y,z}$  integrated over $\theta_z$ and $\theta_y$, respectively. Solid lines: (a) $\bar{I}_{y}(\theta_y; \phi_{\rm CEP})$ and (b) $\bar{I}_{z}(\theta_z; \phi_{\rm CEP})$. Line colors: $\phi_{\rm CEP}=0$ (blue), $\pi/2$ (red), and $\pi$ (yellow). (c) CEP contrasts between $\phi_{\rm CEP}=0$ and $\phi_{\rm CEP}=\pi$:  $\delta \bar{I}_{z}(\theta_z)$ (blue line) and $\delta \bar{I}_{y}(\theta_y)$ (red line). Propagation distance: $ct_1 = 1,000 \lambda_L$. Photons energy: $\hbar \omega_{\rm X-ray} < 3 {\rm keV}$.}\label{fig:CEP_jitter}
\end{figure}

Finally, we analyze the dependence of the angular distribution $I(\theta_y,\theta_z)$ of the beam-generated X-ray flux on the CEP offset $\phi_{\rm CEP}$. If $I(\theta_y,\theta_z)$ is indeed CEP-dependent, then it can be used as a telltale sign of phase-dependent beam dynamics in the plasma bubble. Because multiple electrons undergo several betatron oscillations inside the plasma bubble after their injection, and those electrons may not be fully correlated in their motion, it is not {\it a priori} obvious that the absolute laser phase (characterized by $\phi_{\rm CEP}$) will have a measurable impact on the angular distribution of the emitted X-rays. Broadly speaking, we are attempting to address the following situation: is there an imprint of the absolute phase of a collection of low-energy {\it laser photons} onto a collection of high-energy {\it X-ray photons} generated via a fairly complex and indirect upconversion process of the former? Here $\hbar \omega_{\rm X-ray} \sim 1 {\rm keV}$ while $\hbar \omega_L \sim 0.11 {\rm eV}$, i.e., such up-conversion is inherently an extremely high-order process.

Surprisingly, we find that there is a small but non-negligible CEP dependence on the angular X-ray asymmetry, i.e. $I \equiv I(\theta_y,\theta_z; \phi_{\rm CEP})$ can be expressed as $I = I_0(\theta_y,\theta_z) + \delta I(\theta_y,\theta_z; \phi_{\rm CEP})$, where $I_0$ is CEP-independent even function of $(\theta_y,\theta_z)$, and $|\delta I| \ll I_0$. Not only does the angular X-ray distribution has a significantly larger spread in the $z$-direction than in the $y$-direction (i.e., $I_0(\theta_y,\theta_z)$ is anisotropic and polarization-dependent, as presented in Fig.~\ref{fig:Asymmetry}(a) earlier), but also its small asymmetric deviation from $I_0$ is a function of the laser CEP offset.

To better visualize the dependence of the asymmetry function $\delta I$ on $\phi_{\rm CEP}$, we introduce and plot in Fig.~\ref{fig:CEP_jitter}(a-b) two angle-integrated X-ray fluxes: $\bar{I}_y(\theta_y; \phi_{\rm CEP}) \equiv \int I d\theta_z$ and $\bar{I}_z(\theta_z; \phi_{\rm CEP}) = \int I d\theta_y$. These quantities are plotted in Figs.~\ref{fig:CEP_jitter}(a,b) for three values of $\phi_{\rm CEP} = 0, \pi/2, \pi$. While the dependence of $\bar{I}_y$ on the CEP offset is negligible, there is a small but visible in Fig.~\ref{fig:CEP_jitter}(b) dependence of $\bar{I}_z$ on $\phi_{\rm CEP}$. The corresponding differences between the integrated X-ray fluxes calculated for $\phi_{\rm CEP}=0$ and $\phi_{\rm CEP}=\pi$, defined as the CEP contrasts $\delta \bar{I}_{y}(\theta_y) \equiv \bar{I}_y(\theta_y;0) - \bar{I}_y(\theta_y;\pi)$ and $\delta \bar{I}_{z}(\theta_z) \equiv \bar{I}_z(\theta_z;0) - \bar{I}_z(\theta_z;\pi)$, are plotted in Fig.~\ref{fig:CEP_jitter}(c). The two plots clearly show that $\left| \delta \bar{I}_{z} \right| \gg \left| \delta \bar{I}_{y} \right|$, confirming that there is a much larger CEP contrast for the X-rays emitted in the laser polarization direction.

\subsection{Laser polarization dependence of the spectral brightness and angular distribution of the X-ray flux }\label{subsection:xray_cep}

\begin{figure}[h]
    \includegraphics[width=\textwidth]{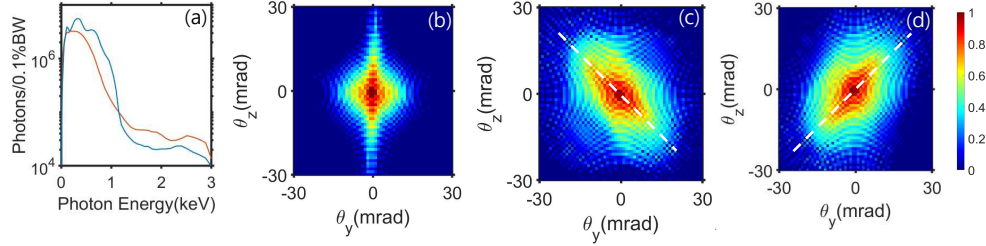}
\caption{CEP dependence of the angular distribution of betatron X-rays for CP and LP laser drivers. Propagation distance: $ct_1 = 1,000 \lambda_L$. Photon energy: $\hbar \omega_{\rm X-ray} < 3 {\rm keV}$. (a) X-ray spectra for LP (red) and CP (blue) laser pulses. (b-d) X-ray intensity distribution for photons below 3keV generated by LP driver $\phi_{\rm CEP}=0$(b) , CP driver $\phi_{\rm CEP}=0$ (c) and CP driver, $\phi_{\rm CEP}=\pi/2$(d). White lines: 135 degree(c), 45 degree(d). Parameters: same as in Fig.~\ref{fig:Asymmetry}. Propagation distance: $ct_1 = 1,000 \lambda_L$. Photons energy: $\hbar \omega_{\rm X-ray} < 3 {\rm keV}$.} \label{fig:Polarization}
\end{figure}

As noted earlier, different polarizations of the driving laser pulse result in distinct phase space distributions of the accelerated electrons:  see Figs.~\ref{fig:spectra}(a-b) for the comparison between the CP and LP cases. Consequently, the radiated photon spectra are also expected to vary accordingly. For example, a LWFA driven by a CP pulse generates more photons at the lower energy range, but an LP-driven LWFA generates more photons at higher energies as shown in  Fig.~\ref{fig:Polarization}(a). Likewise, the angular distribution of the X-ray flux produced by the betatron radiation of the accelerated electrons is also polarization-dependent. Confirming our findings from the swarm simulations (see Figs.~\ref{fig:test_radiation}(c-g)), an LP driver generates X-ray with angular distribution that is strongly elongated in the laser polarization direction, as shown in Fig.~\ref{fig:Polarization}(b). Notably, the elongation direction does not depend on the CEP offset, as was shown earlier in Figs.~\ref{fig:CEP_jitter}(a,b).

While a CP laser driver also generates angularly-anisotropic X-ray flux elongated in one direction, the elongation direction itself can, in principle, be dependent on the CEP offset of the laser pulse. The CEP effect on the angular X-ray distribution $I(\theta_y,\theta_z; \phi_{\rm CEP})$ in the CP case is illustrated in Figs.~\ref{fig:Polarization}(c-d), where the intensity ellipse is rotated from the $-45^{\circ}$ position for $\phi_{\rm CEP} = 0$ to $+45^{\circ}$ position for $\phi_{\rm CEP} = \pi/2$. This is because the angle at which highest energy electrons are injected rotates according to the CEP offset of the CP pulse, leading to most intense betatron radiation generated at the same angle.

Such asymmetric CEP-controlled distribution of X-rays occurs despite the fact that the electron injection is mostly continuous in the case of CP driver. Nevertheless, injected electrons ``remember" the direction of the laser pulse at the time of laser pulse steepening. This direction, which is determined by the CEP offset of a CP-polarized laser, is reflected by the direction of a helical transverse wake ${\bf W}_{\perp}$ at the back of the bubble, where electron injection takes place. Characterization of the phase of CP laser pulses has been proposed in the case of ultra-intense lasers ~\cite {CEP_CP_detection}, but it remains an open area of active research for multi terawatt class lasers. We envision that betatron radiation may enable the characterization of CEP offsets of CP laser pulses.

\section{Conclusions}

In this paper, we proposed the concept of an Expanding Phase-controlled Undulating Bubble (EPUB), in which a self-steepened few-cycle multi-TW laser pulse produces an expanding and undulating plasma bubble in its wake. EPUB can trap electron bunch with spatiotemporal structure. The degree of bunching is controlled via changing laser polarization, alternating between highly modulated high-current beam or a flat-current beam with fs scale modulation.

PIC simulation shows that indeed a structured beam with larger charge(O(nC)) is trapped. Appreciable fraction of the injected beam forms a highly mono energetic energy peak, and a large fraction (50\%) of laser pulse energy is efficiently transferred  to the injected electrons. The modulation period is altered via changing laser power or density, and the precise location of some of the injected bunch can be controlled via changing laser CEP. Betatron radiation from the trapped electrons shows distinct intensity and angular distribution with clear dependence on laser polarization and CEP.  By measuring electron beam pointing or X-ray intensity distribution at this stage, the absolute CEP of the laser can be retrieved.

Our injection and acceleration scheme is not limited to mid-infrared (e.g., CO$_2$) lasers, and can be applied to shorter wavelength few-cycle TW class lasers. This will provide an efficient way to generate bright betatron X-ray source in the soft X-ray regime switching between pulsed and continuous configuration with controllable duration and time-delay. It can also serve as a way to calibrate or retrieve the CEP of any laser system that does not have phase stabilization. High-flux polarized X-ray radiation (including soft X-rays with energies below $\hbar \omega_{\rm X-ray} < 1 {\rm keV}$) is of great importance for element-specific studies in a variety of scientific fields, including wet cell biology~\cite{kleine_JPC19}, condensed matter physics, extreme ultraviolet optics technology, and warm dense matter~\cite{kleine_JPC19,mahieu_ncomm18,dopp_Optica18,wenz_ncomm15}. Of particular interest for time-resolved X-ray absorption spectroscopy (XAS) are femtosecond broadband features~\cite{mahieu_ncomm18} of betatron radiation. It is likely that linearly-polarized femtosecond broadband X-rays obtained using EPUB-based laser-plasma acceleration will be of interest to these and other applications.

 \section{Acknowledgments}
This work was supported by the Department of Energy under a Grant No. DE-SC0019431 and by the National Science Foundation under a Grant No. PHY-2109087. The authors thank Dr. Roopendra Rajawat for helpful discussions and feedbacks and the Texas Advanced Computing Center (TACC) at The University of Texas at Austin for providing the HPC resources.

\appendix

\section{Control of the injected/accelerated bunch profile using laser intensity and carrier-envelope phase offset}\label{section:bunch control}

In this Appendix, we describe additional PIC simulations demonstrating that the current profile of the injected and accelerated electron bunch can be further controlled by the intensity and the CEP of the laser pulse. The same spatial/temporal laser parameters as in Section~\ref{sec:pic} (see the caption of Fig.~\ref{fig:sim} and Table ~\ref{table1}) and two laser powers (low: $P_L^{(1)} = 40 {\rm TW}$ and high: $P_L^{(2)} = 50 {\rm TW}$) are used.


\begin{figure}[h]
    \includegraphics[width=\textwidth]{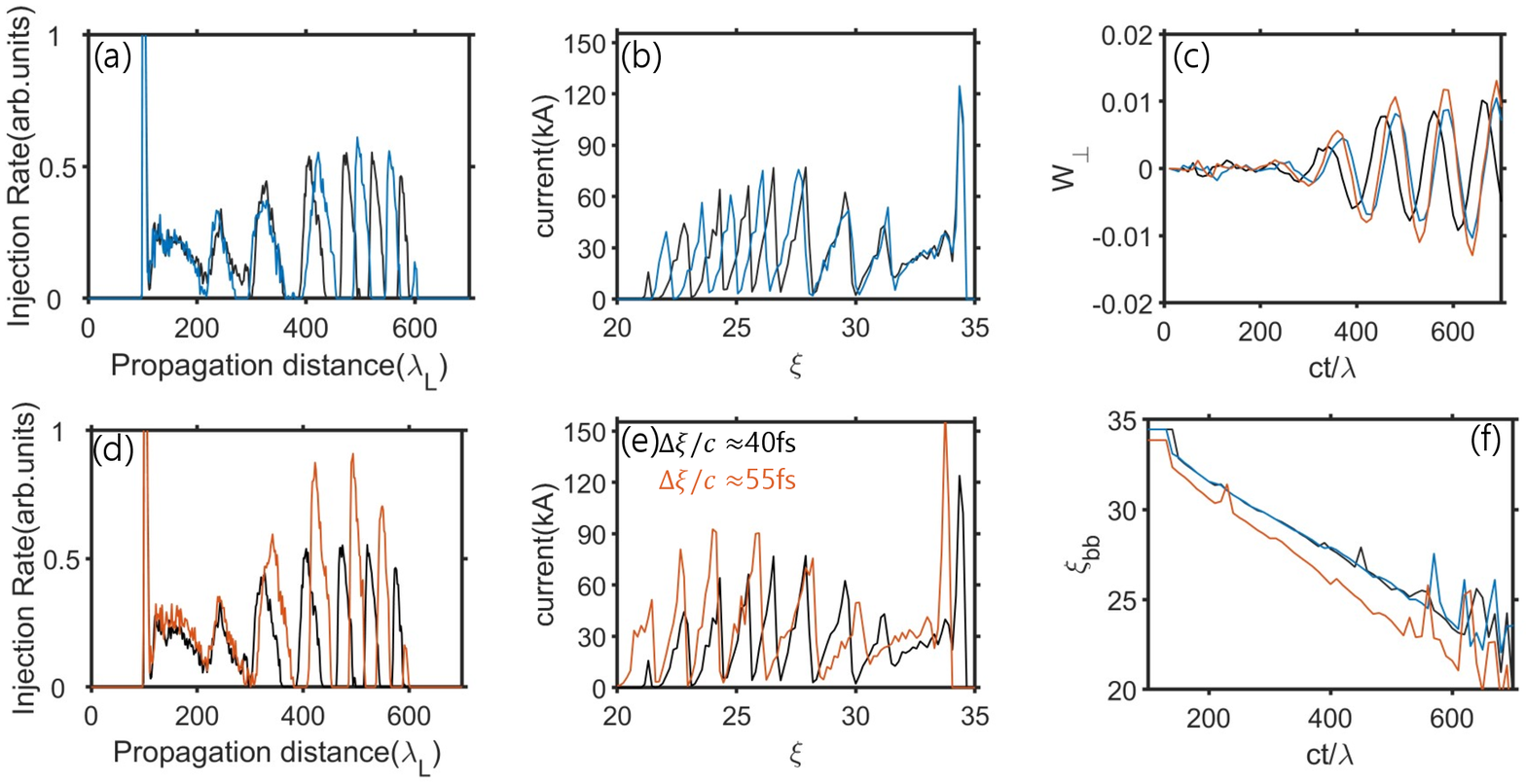}
\caption{
Injection control using LP laser CEP and power. (a-b) Injection rate(a) and current(b) for different CEP with same laser power(40TW). (d-e) Injection rate(d) and current(e) for different power with same CEP. Transverse wake at $\xi=35$ (c), and bubble rear position(f). 40TW $\phi_{\rm CEP}=0$ laser pulse(black line), 40TW, $\phi_{\rm CEP}=\pi/2$ laser pulse(blue line), 50TW, $\phi_{\rm CEP}=0$ laser pulse (orange line) }\label{fig:powerdep}
\end{figure}

While bunches are injected into the bubble throughout multiple undulation cycles, the absolute value of the CEP offset $\phi_{\rm CEP}$ of a few-cycle pulse described in Sec.~\ref{sec:pic} ($cT_{\rm FWHM}/\lambda_L \sim 3$) can have direct effect on the temporal profile of the injected electron bunch. To analyze the dependence of the current profile on $\phi_{\rm CEP}$, we carried out PIC simulations for the low-power case ($P_L = P_L^{(1)}$) laser pulse with $\phi_{\rm CEP}=0$ and $\phi_{\rm CEP}=\pi/2$. Because $\phi_{\rm CEP}$ controls the phase of the bubble undulation $z_{\rm osc}(t)$ in the laser polarization direction $z$, and that of the transverse wake $W_{z}(t,\xi)$, electron injection times are also $\phi_{\rm CEP}$-dependent.  This can be observed in Figs.~\ref{fig:powerdep}(a,b), where the injection rates and the current profiles are plotted as a black ($\phi_{\rm CEP}=0$) and blue ($\phi_{\rm CEP}=\pi/2$) lines in Figs.~\ref{fig:powerdep}(a) and (b), respectively.

Not surprisingly, only Group II and III electrons (see Sec.~\ref{subsec:electron_groups} for the definitions of the three electron groups) are affected by the CEP offset because bubble undulations can either suppress or enhance electron injections in a phase-dependent manner. Namely, the locations of the troughs in the $\phi_{\rm CEP}=0$ case become locations of peaks for $\phi_{\rm CEP}=\pi/2$ case. On the other hand, Group I (i.e. earlier injected electrons) are not significantly controlled by the absolute value of $\phi_{\rm CEP}=\pi/2$ because (i) their undulation amplitude is small, and (ii) bubble undulations neither suppress not enhance electron injections. Other CEP-independent effects, such as beam loading, play a more dominant role in determining injection dynamics. For completeness, the amplitude of the transverse wake $W_{z}(t,\xi_0)$ at the fixed position $\xi_0=35$ near the rear of the bubble is plotted in Fig.~\ref{fig:powerdep}(c) for the two CEP phases as a black line ($\phi_{\rm CEP}=0$) and a blue line ($\phi_{\rm CEP}=\pi/2$). The two curves are shifted in time by $\pi/2$ according to Fig.~\ref{fig:powerdep}(c).

In addition, the transverse wake is plotted in Fig.~\ref{fig:powerdep}(c) for a more intense laser pulse with the peak power $P_L = P_L^{(2)}$. As expected, the transverse wake (orange line) exceed in magnitude that produced by the lower-intensity laser with $P_L = P_L^{(2)}$ (black and blue lines). However, the oscillation period $T_{\rm CEP}$ of the transverse wake is the same for both laser powers. This is expected because the expression for $T_{\rm CEP}$ is intensity-independent: $T_{\rm CEP} \approx \lambda_L/{(v_{\rm ph}-v_{\rm g})} \sim (\lambda_L/c) (\omega_L^2/\omega_p^2)$. Therefore, the injection period is also intensity-independent as can be seen in Fig.~\ref{fig:powerdep}(d), where the injection rates along the laser propagation direction $x$ are plotted for $P_L = P_L^{(1)}$ (black line) and $P_L = P_L^{(2)}$ (orange line); the same $\phi_{\rm CEP} = 0$ was used in both simulations.

The modulation period $\Delta \xi$ of the injected electrons is, however, highly sensitive to the velocity $v_{\rm bb}$ of the back of the expanding plasma bubble, which in turn depends on the rate of plasma bubble expansion according to $v_{\rm bb}/c = v_{\rm b}/c - k_p R_b \varepsilon$. As shown in Sec.~\ref{subsec:electron_groups}, the resulting compression effect of the injected electron bunches results in their periodicity given by $\Delta \xi \approx cT_{\rm CEP}/4\gamma_{\rm bb}^2$, where $\gamma_{\rm bb}\approx 1/\sqrt{1-v_{\rm bb}^2/c^2}$. These dependencies suggest that, by keeping plasma density and laser frequency same but increasing the pulse power, one can increase the bubble radius $R_b$ and the expansion rate $\varepsilon$. This would effectively slow down the back of the plasma bubble and result in a smaller $\gamma_{\rm bb}$, leading to reduced bunch compression and increased current modulation period $\Delta \xi$.

By plotting the location $\xi_{\rm bb}$ of the back of the plasma bubble for the two laser powers in Fig.~\ref{fig:powerdep}(f), we indeed confirm that $v_{\rm bb}$ becomes slower for $P_L = P_L^{(2)}$ (orange line) that for $P_L = P_L^{(1)}$ (black line). Quantitatively, $\gamma_{\rm bb}^{(1)} \approx 5$ and $\gamma_{\rm bb}^{(2)} \approx 4.5$. This results in a longer bunch modulation period for the higher-power laser pulse: we observe from Fig.~\ref{fig:powerdep}(e) that the $20<\xi<33$ window contains $8$ current peaks for the $P_L = P_L^{(1)}$ case (corresponding to $\Delta \xi^{(1)}/c \approx 40 {\rm fs}$), and only $6$ current peaks for the $P_L = P_L^{(2)}$ case (corresponding to $\Delta \xi^{(2)}/c \approx 55 {\rm fs}$). Therefore, both the period and the absolute timing of laser-accelerated femtosecond electron bunches can be controlled using the CEP offset and the peak power of a few-cycle laser pulse.

\section{ Equation of motions from the Moving Frame Hamiltonian}\label{Hamiltonian}
In this Appendix, we introduce the equations of motion for the electron interacting with a moving bubble which we use throughout the text. 

From the moving frame Hamiltonian in Section ~\ref{sec:analytics}, the following equations of motion can be derived using $d\mathbf{P}/dt=-\partial H/\partial\boldsymbol{\rho}$, $d \boldsymbol{\rho}/dt=\partial H/\partial \mathbf{P}$:
\begin{eqnarray}
\label{eq:eqn1}
&\frac{d\xi}{dt} = \frac{p_x}{\gamma} - v_{\rm b}, \\ \label{eq:eqn2}
&\frac{d p_x}{dt} =-\frac{1}{4} \left[ R(t) \dot{R}(t)+\xi(1+v_{\rm b}) + (v_z - \dot{z}_{\rm osc}) \tilde{z}\right],&  \\
\label{eq:eqn3}
&\frac{dz}{dt} = \frac{p_z}{\gamma},  \\
\label{eq:eqn4}
&\frac{d p_{z}}{dt} = -\frac{(v_x+1)\tilde{z}}{4}, \\
\end{eqnarray} which are used to numerically compute the electron trajectories in the test-particle simulations.

A convenient way to describe electron interaction with a non-evolving plasma bubble has been derived~\cite{kost_injection} under which time, length, and momentum are normalized ($s=t/R,X=\xi/R, Z=z/R, P_x=p_x/R^2,P_z=p_z/R^2$). Dropping terms of $O(R^{-4}), O(\gamma_{\rm b}^{-2})$ from equations of motion similar to ~\ref{eq:eqn1}-\ref{eq:eqn4} for a non-evolving bubble, differential equations independent of bubble radius can be found, which we reproduce here for convenience.

\begin{eqnarray}
\label{eq:eqn6}
&\frac{d P_x}{ds} = -\frac{X}{2}+ \frac{P_z}{P_x^2+P_y^2} \\
\label{eq:eqn7}
&\frac{d P_z}{ds} =-\frac{Y}{4} \left(1+\frac{P_x}{\sqrt{P_x^2+P_z^2}} \right),&  \\
\label{eq:eqn8}
&\frac{dX}{ds} = \frac{P_x}{\sqrt{P_x^2+P_z^2}}-1,  \\
\label{eq:eqn9}
&\frac{dY}{ds} = \frac{P_y}{\sqrt{P_x^2+P_z^2}}.
\end{eqnarray}

These equations can be numerically solved to find electron momentum and position by re-scaling to physical variables. For a slowly evolving bubble, these equations are convenient ways to estimate the zeroth order quantities, as was done in Section ~\ref{sec:analytics}.

\end{document}